\documentclass[iop]{emulateapj}

\usepackage{epsfig, epstopdf, graphicx, color}
\usepackage{enumitem}
\usepackage{ulem}
\usepackage{float} 
\usepackage{amssymb}
\usepackage{amsfonts}
\usepackage{amsbsy}
\usepackage{amsmath}
\usepackage{natbib}
\usepackage{mathrsfs}
\usepackage{lineno}
\usepackage{eso-pic}
\usepackage{hyperref} 
\usepackage[bottom]{footmisc}
\usepackage{footnote}
\usepackage{fancyheadings}

\def \lymanalpha {Ly$\alpha$}
\def \OII 		 {[OII]3727}
\def \OIII       {[OIII]5007}
\def \Hbeta		 {{\rm H}_{\beta}}

\def \zsource     {z_{\rm spec}}

\def \zlens       {z_{\rm l}}
\def \zcluster    {z_{\rm c}}
\def \zredmapper  {z_{\rm RM}}
\def \thetae      {\theta_{\rm E}}
\def \menclosed   {M_{\rm enc}}

\def \kms  {{\rm km\, s^{-1}}}
\def \msol {{\rm M_{\odot}}}
\def \sqdeg {sq. deg.\ } 
\def \sqdegadjective {-sq.-deg.\ }
\def \sqarcsec {{\rm arcsec}^2}
\def \massunitshi {\times 10^{13}}
\def \massunitslo {\times 10^{12}}
\def \arcsec {''}

\def \opensearch {non-targeted} 

\def \mysim {=}

\def \SystemADES {DES J0221-0646} 
\def \SystemBDES {DES J0250-0008} 
\def \SystemCDES {DES J0329-2820} 
\def \SystemDDES {DES J0330-5228} 
\def \SystemEDES {DES J0446-5126} 
\def \SystemFDES {DES J2336-5352} 

\def \arcA {A}
\def \arcB {B}
\def \arcC {C}

\def \redmapper {redMAPPer}

\def \SystemAspeczA {2.7251\pm0.0008}
\def \SystemAspeczB {2.7241\pm0.0008}
\def \SystemBspeczA {1.2081\pm0.0004}
\def \SystemCspeczA {0.7963\pm0.0001}
\def \SystemCspeczB {1.2976\pm0.0003}
\def \SystemDspeczA {1.4541\pm0.0004}
\def \SystemEspeczC {3.2086\pm0.0011}
\def \SystemEspeczB {3.2068\pm0.0010}
\def \SystemEspeczA {0.1680\pm0.0009}
\def \SystemFspeczA {1.1528\pm0.0006}
\def \SystemFspeczB {0.8972\pm0.0004}

\def \SystemAradiusA {5.0\pm1.4}
\def \SystemBradiusA {6.6\pm1.1}
\def \SystemCradiusA {7.2\pm1.4}

\def \SystemDradiusA {6.1\pm1.5}
\def \SystemEradiusB {7.0\pm1.5}
\def \SystemFradiusA {5.0\pm1.5}
\def \SystemFradiusB {8.6\pm1.9}

\def \SystemAmencA {7.5\pm4.7} 
\def \SystemBmencA {3.7\pm3.0} 
\def \SystemCmencA {1.6\pm0.9} 
\def \SystemDmencA {9.0\pm3.7} 
\def \SystemEmencB {1.6\pm0.9} 
\def \SystemFmencA {8.6\pm7.7} 
\def \SystemFmencB {3.5\pm3f3} 

\def \SystemAmencAtext {\SystemAmencA \massunitslo\, \msol}
\def \SystemBmencAtext {\SystemBmencA \massunitshi\, \msol}
\def \SystemCmencAtext {\SystemCmencA \massunitshi\, \msol}

\def \SystemDmencAtext {\SystemDmencA \massunitslo\, \msol}
\def \SystemEmencBtext {\SystemEmencB \massunitshi\, \msol}
\def \SystemFmencAtext {\SystemFmencA \massunitslo\, \msol}
\def \SystemFmencBtext {\SystemFmencB \massunitshi\, \msol}

\def \SystemAmencAtable {\SystemAmencA \massunitslo}
\def \SystemBmencAtable {\SystemBmencA \massunitshi}
\def \SystemCmencAtable {\SystemCmencA \massunitshi}

\def \SystemDmencAtable {\SystemDmencA \massunitslo}
\def \SystemEmencBtable {\SystemEmencB \massunitshi}
\def \SystemFmencAtable {\SystemFmencA \massunitslo}
\def \SystemFmencBtable {\SystemFmencB \massunitshi}

\def \slitidname {Source Image ID}

\def \figurescalesystem {0.8}
\def \figurescalemultipanel {0.9}

\def \figuremultipanel 
{
\begin{figure*}[!htb]
\centering
\includegraphics[scale=\figurescalemultipanel]{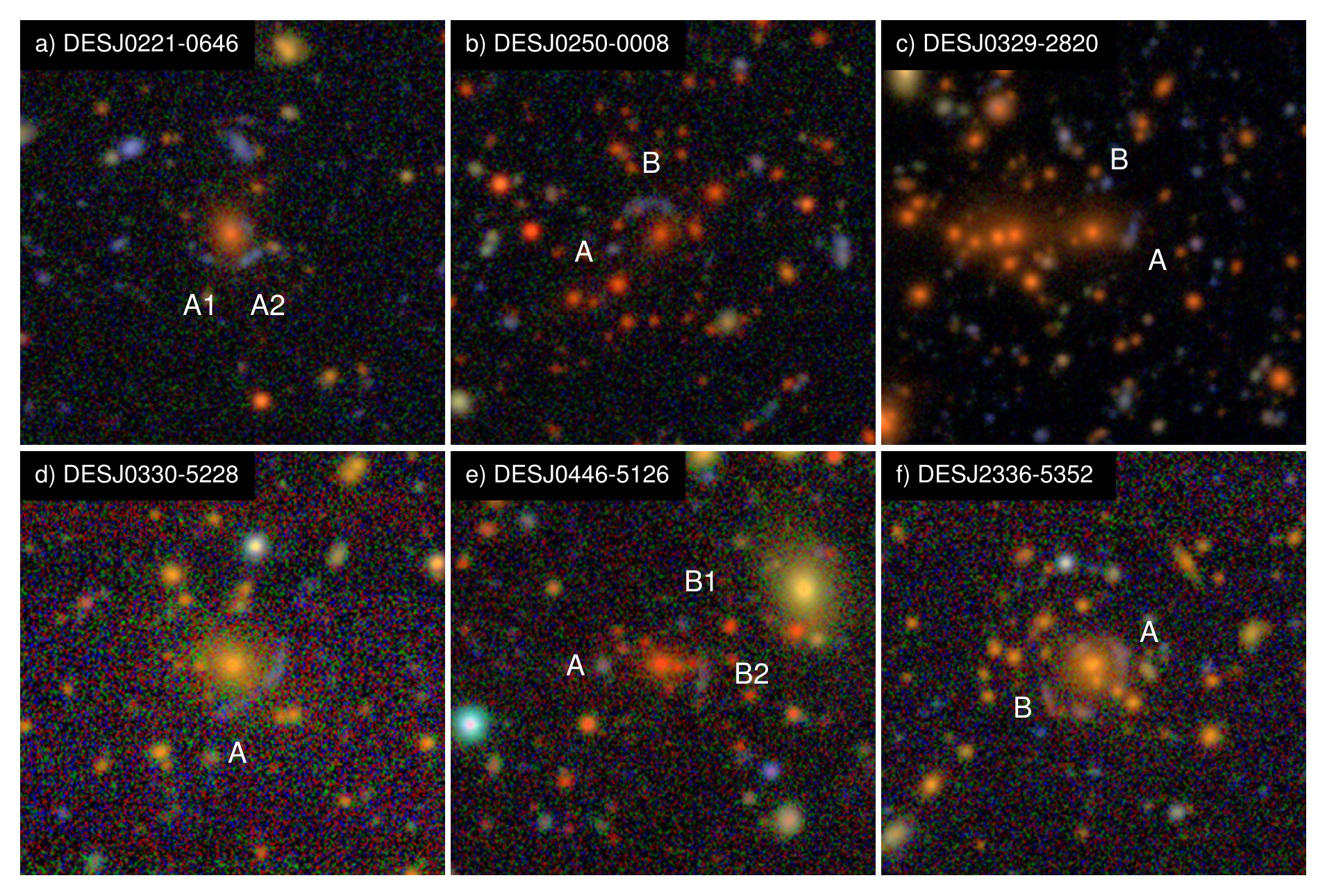}
\caption{Color co-added DES images of the six systems described in this work: a) \SystemADES, b) \SystemBDES, c) \SystemCDES, d) \SystemDDES, e) \SystemEDES, f) \SystemFDES. All images are oriented north up, east left, and are $1'\times 1'$ in area. The lensing features targeted for spectroscopy are indicated by the letters. refer to Fig.'s~\ref{fig:SystemA03}-\ref{fig:SystemF24} for a more detailed view of the field and the image candidates}
\label{fig:multipanel}
\end{figure*}
}

\def \FigureSystemA
{
\begin{figure*}[!htb]
\centering
\includegraphics[scale=\figurescalesystem]{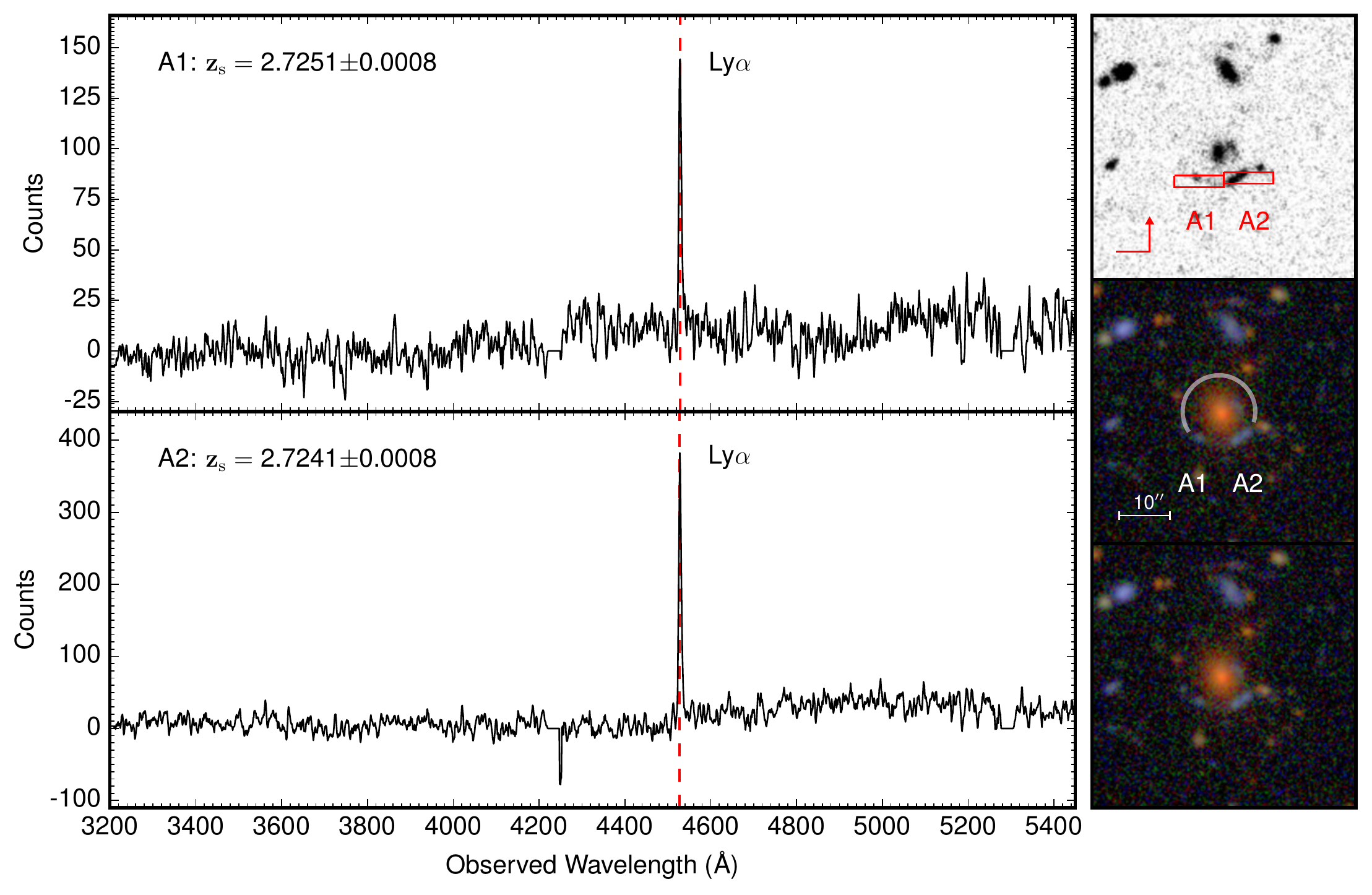}
\caption{\SystemADES. The 1D spectra for A1 and A2 are shown on the left. There are strong  emission lines at the same observed wavelength $4535$\AA\ in both spectra. Taking into account the absence of other spectral features in the R150 spectra, along with the photometric redshift of the lens galaxy, we assign the features to be \lymanalpha, which gives redshifts of $\zsource=\SystemAspeczA$ and $\SystemAspeczB$ for A1 and A2, respectively. The orientation of the slits (red boxes) is shown in the upper right DES $g$-band image. In the color coadd image in the middle right panel, the two features of interest are labeled ``A1'' and ``A2'', the Einstein radius is marked by a white arc, and the small scale bar shows the size of the image. In the lower right is an identical color coadd image. All images are oriented north up, east left. The flat, zero-count portions of the spectra near $\lambda \sim 4200$\AA\ and $\sim5300$\AA\ show the chip gaps in the detectors: we didn't take an offset observation to cover the chip gaps for this system.}
\label{fig:SystemA03}
\end{figure*}
}

\def \FigureSystemB
{
\begin{figure*}[!htb]
\centering
\includegraphics[scale=\figurescalesystem]{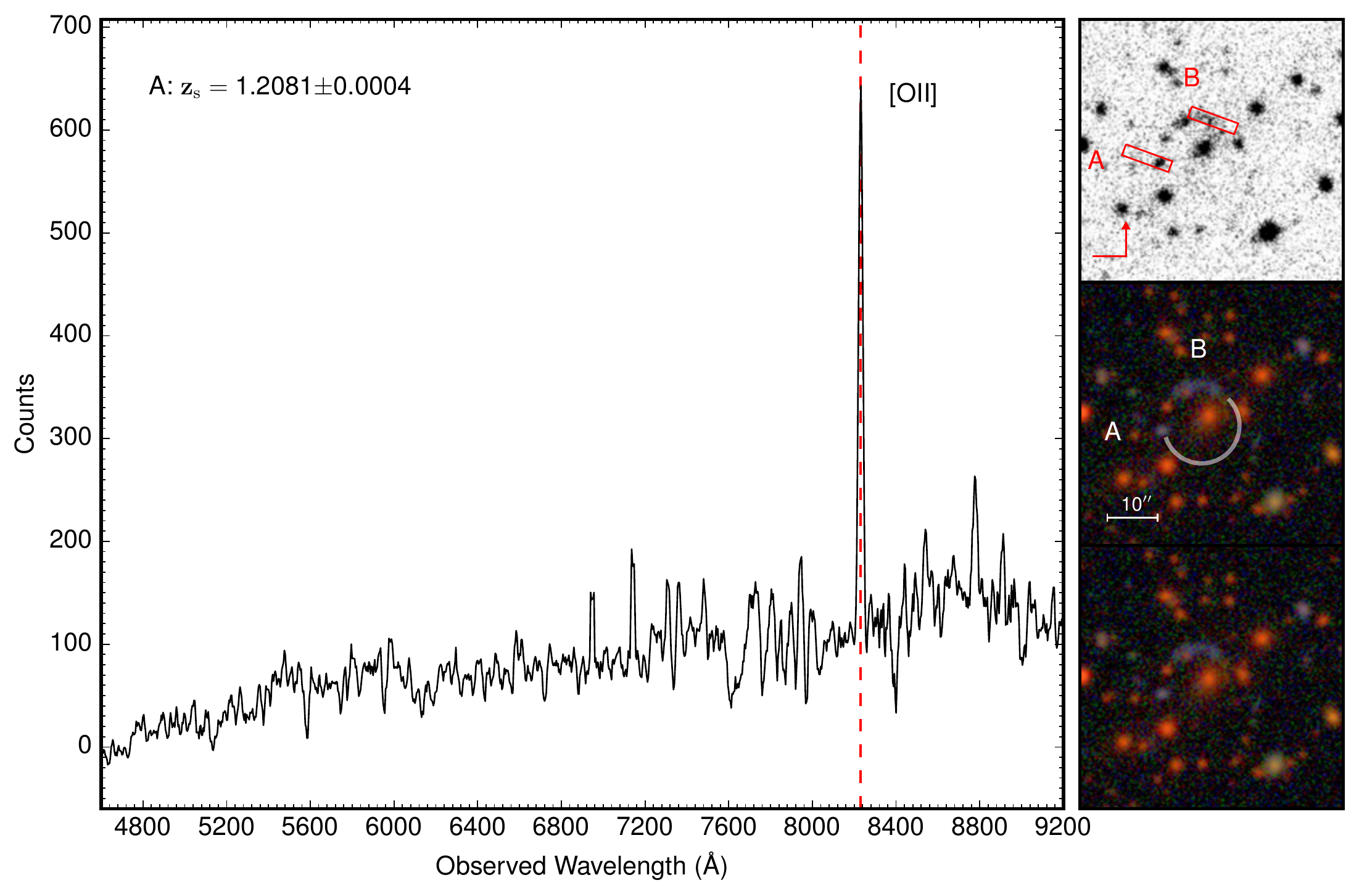}
\caption{\SystemBDES. The 1D spectrum for \arcA\ is shown on the left. There is a strong emission line at an observed wavelength of $8229$\AA. We associate this feature with \OII, which yields a redshift of $\zsource=\SystemBspeczA$. The orientation of the slits (red boxes) is shown in the upper right DES $r$-band image. In the color coadd image in the middle right panel, the two features of interest are labeled ``\arcA'' and ``\arcB'', the Einstein radius is marked by a white arc, and small scale bar shows the size of the image. In the lower right is an identical color coadd image. All images are oriented north up, east left.}
\label{fig:SystemB05}
\end{figure*}
}

\def \FigureSystemC
{
\begin{figure*}[!htb]
\centering
\includegraphics[scale=\figurescalesystem]{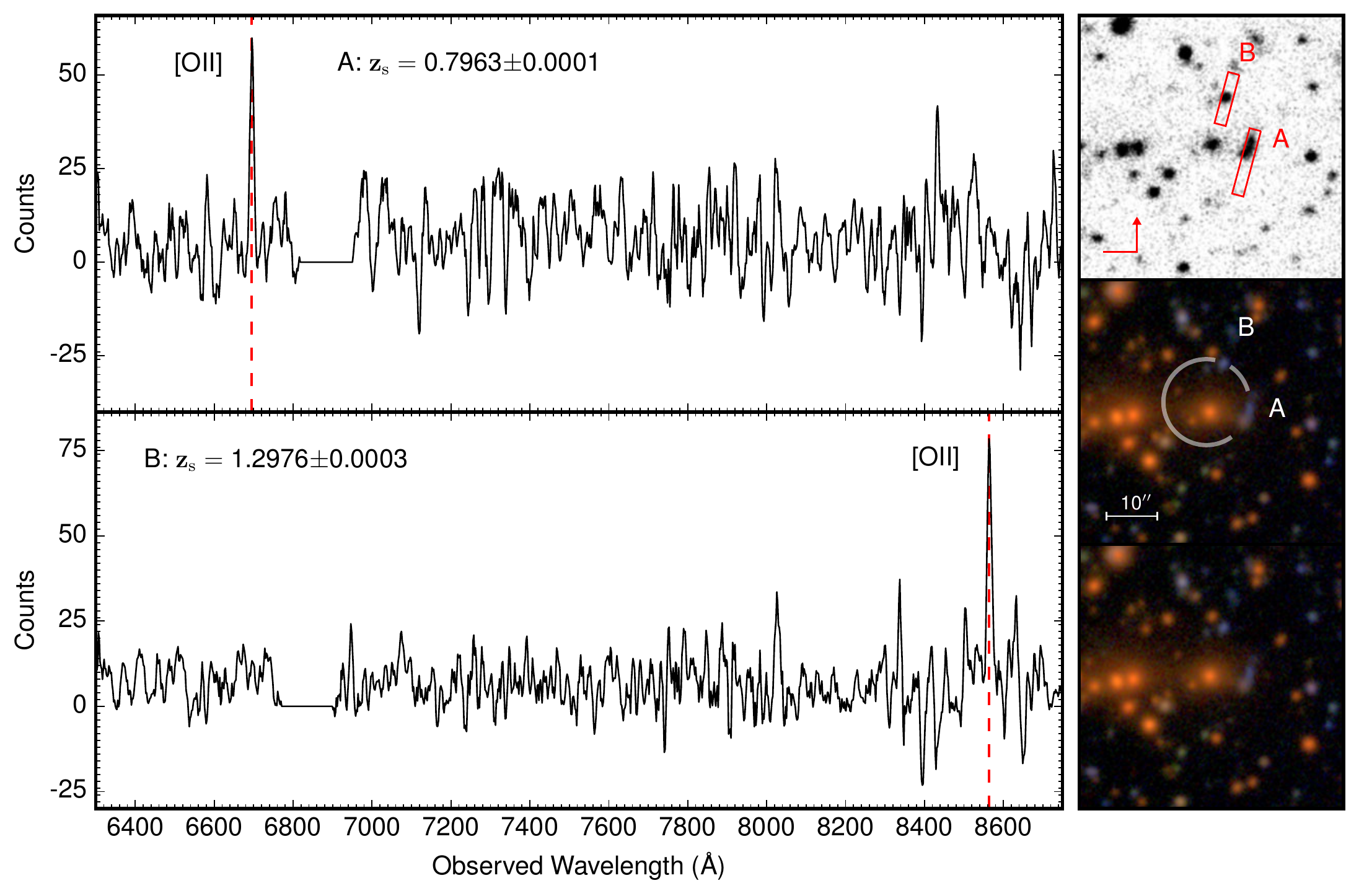}
\caption{\SystemCDES. The 1D spectra for \arcA\ and \arcB\ are shown on the left. There are emission lines at two different observed wavelengths, 6694\AA\ and 8563\AA. Given the lack of observed features at shorter wavelengths, we identify the emission as \OII, from which we obtain redshifts, $\zsource=\SystemCspeczA$ and $\SystemCspeczB$ for \arcA\ and \arcB, respectively. The orientation of the slits (red boxes) is shown in the upper right DES $g$-band image. In the color coadd image in the middle right panel, the two features of interest are labeled ``\arcA'' and ``\arcB''. The Einstein radii are marked by white and red arcs for slits \arcA\ and \arcB, respectively.  The small scale bar shows the size of the image. In the lower right is an identical color coadd image. All images are oriented north up, east left. The flat, zero-count portions of the spectra near $\lambda \sim 6800$\AA\ show the chip gaps in the detectors: we didn't take an offset observation to cover the chip gaps for this system.}
\label{fig:SystemC06}
\end{figure*}
}

\def \FigureSystemD 
{
\begin{figure*}[!htb]
\centering
\includegraphics[scale=\figurescalesystem]{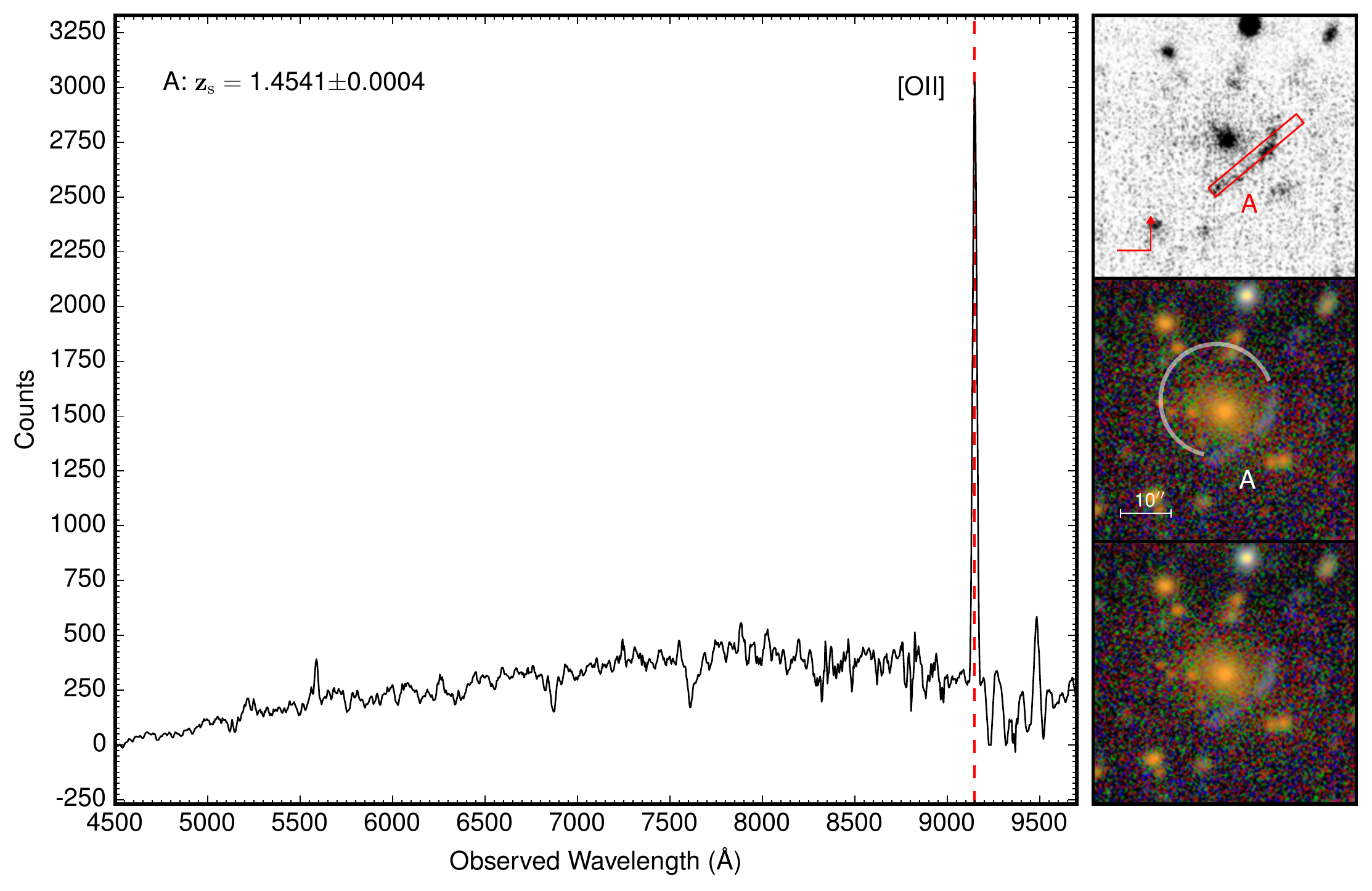}
\caption{\SystemDDES. The 1D spectrum for \arcA\ is shown on the left. There is a strong emission line at an observed wavelength of 9146\AA, which we take to be \OII, from which we obtain a redshift of $\zsource=\SystemDspeczA$. The orientation of the slit (red boxes) is shown in the upper right DES $g$-band image. In the color coadd image in the middle right panel, the feature of interest is labeled ``\arcA'', the Einstein radius is marked by a white arc, and a small scale bar shows the size of the image. In the lower right is an identical color coadd image. All images are oriented north up, east left.}
\label{fig:SystemD07}
\end{figure*}
}

\def \FiguresystemE
{
\begin{figure*}[!htb]
	\centering
	\includegraphics[scale=\figurescalesystem]{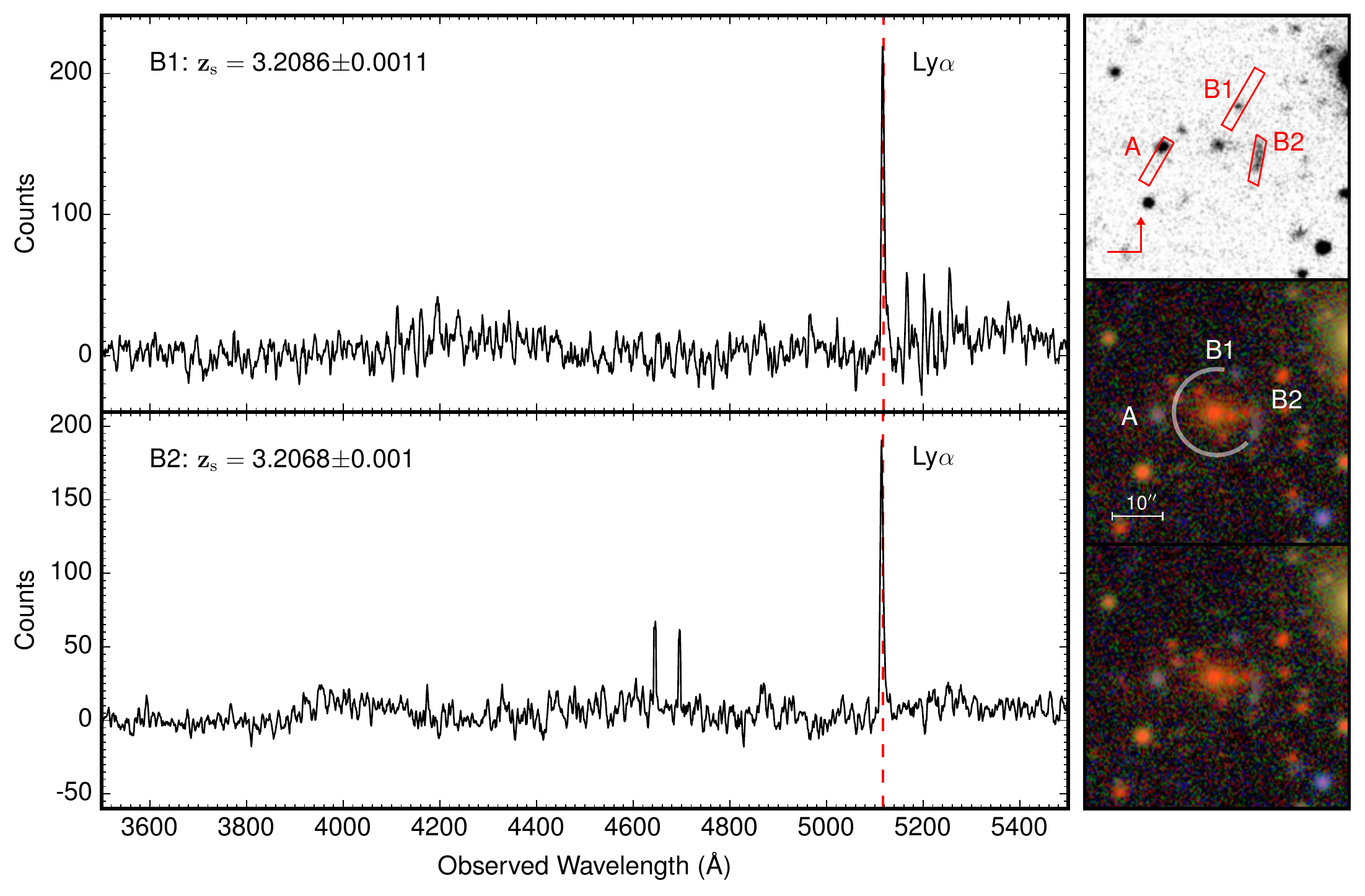}
	\caption{\SystemEDES. The 1D spectra for B1 and B2 are shown on the left. The spectrum for \arcA\ is not shown, but has an emission line consistent with an interpretation as a foreground object. Prominent emission lines in both arcs at the same observed wavelength of 5117\AA. We conclude that this is \lymanalpha\ emission, from which we obtain redshifts of $\zsource=\SystemEspeczB$ for \arcB\ and $\SystemEspeczC$ for \arcC. The object labeled \arcA\ also has emission lines (not shown), which we interpret as \OII\ and \OIII. This gives a redshift of $\zsource = \SystemEspeczA$, making object \arcA\ a foreground object. The orientation of the slits (red boxes) is shown in the upper right DES $g$-band image. In the color coadd image in the middle right panel the three features of interest are labeled ``A'', ``B1'' and ``B2'', the Einstein radius is marked by a white arc, and the small scale bar shows the size of the image. In the lower right is an identical color coadd image. All images are oriented north up, east left.}
	\label{fig:SystemE14}
\end{figure*}
}

\def \FigureSystemF
{
\begin{figure*}[!htb]
\centering
\includegraphics[scale=\figurescalesystem]{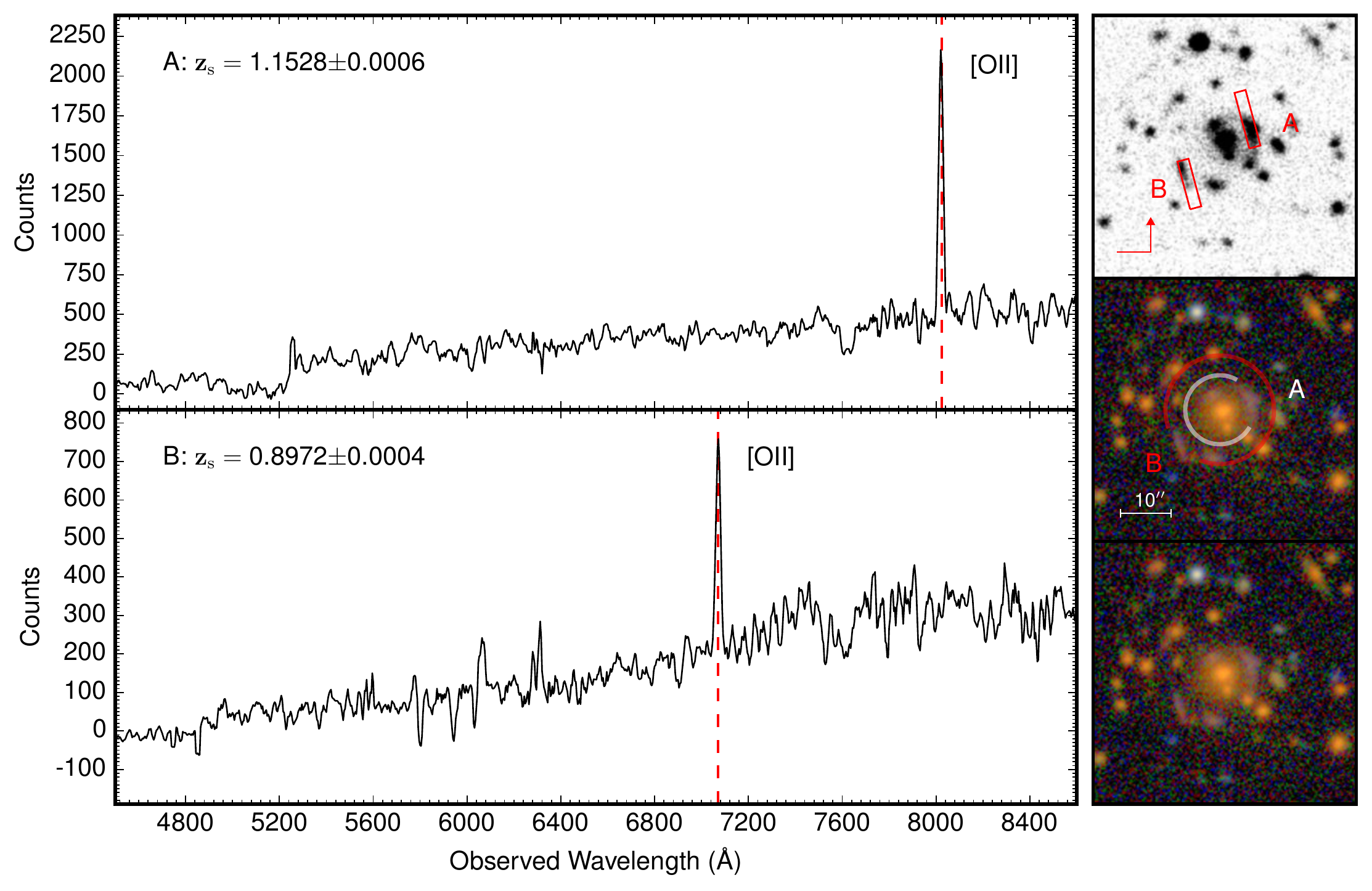}
\caption{\SystemFDES The 1D spectra for A and B are shown on the left. The prominent emission lines at two different observed wavelengths of 8023\AA\ for \arcA\ and 7070\AA\ for \arcB\ are due to \OII. These lines yield source spectroscopic redshifts of $\zsource=\SystemFspeczA$ and $\SystemFspeczB$ \arcA\ and \arcB, respectively. The orientation of the slits (red boxes) is shown in the upper right DES $g$-band image. In the color coadd image in the middle right panel the two features of interest are labeled ``\arcA'' and ``\arcB''. The Einstein radii are marked by white and red arcs for slits \arcA\ and \arcB, respectively. The small scale bar shows the size of the image. In the lower right is an identical color coadd image. All images are oriented north up, east left.}
\label{fig:SystemF24}
\end{figure*}
}

\def \tableobservationlog
{
\begin{table*}[!htb]
  \caption{Spectroscopic Observation Log  \label{table:observationlog}} 
  \centering 
  \footnotesize
  \begin{tabular}{ c c c c c c } 
    \hline\hline             
    DES System ID &  UT Date       & Telescope-     & Grating 	& Total Integration	& Seeing 	 \\ 
                      &                & Instrument     &         	&     (hours)    	&  ($\arcsec$)  	 \\ [0.5ex] 
\hline  \hline    
    \SystemADES       & 2014 Oct 10,19 & Gemini-GMOS    & B600/R150 & 1/3.7		 		& 0.92/0.86  \\
    \SystemBDES 	  & 2014 Oct 24    & Gemini-GMOS    & B600/R150 & 1/1  	  	 		& 0.67/0.66  \\
    \SystemCDES 	  & 2014 Oct 23    & Magellan-IMACS & 200 $\ell$/nm  &    1 	 			& 1.0   	 \\
    \SystemDDES 	  & 2014 Oct 22    & Gemini-GMOS    & R150    	& 1		 			& 0.71  \\
    \SystemEDES 	  & 2014 Oct 19,23 & Gemini-GMOS    & B600/R150	& 1/1.63		 	& 0.64/0.67  \\
    \SystemFDES 	  & 2014 Oct 21    & Gemini-GMOS    & R150    	& 1  	  	 		& 0.73  \\
 [1ex]
 \hline \hline
\end{tabular}\\
\begin{flushleft}
{\bf Notes. } Observation log for follow-up spectroscopic observations. DES ID is  derived from the RA and Dec. position of the lensing object at the center of the system.  For the Gemini observations, seeing is taken from the Gemini Observation Logs, which keep environmental conditions in two-hour increments: the seeing for each exposure set is interpolated for the time of observation between adjacent recordings of the environmental log. For the Magellan observation, seeing is measured from the full-width at half-max of the acquisition star PSFs in the science exposures.
\end{flushleft}
\end{table*}
\normalsize
}

\def \tablelensobjects
{
\begin{table*}[!htb]
  \caption{Properties of Lensing Systems: Positions and Photometry of Lenses and Source Images \label{table:lensingobjects}} 
  \centering 
  \footnotesize
  \begin{tabular}{ c c c c c c c } 
    \hline\hline             
	System ID and  & Object ID  	& RA         	& Dec.  			& (g,r,i,z,Y)                                               \\ 
	\slitidname &            	& (deg)      	& (deg)      	&                                                         \\ [0.5ex]                    
\hline  \hline    
 \SystemADES 	& 2937883420 	& 35.463205 	& -6.792405  	& ($23.32\pm0.03,\,21.79\pm0.01,\,20.73\pm0.00,\,20.25\pm0.00$)\,\ \ \ \ \ \ \ \ \ \ \ \ \ \ \ \ \ \ \  \\
 A1       	 	& 2937882680 	& 35.464063  	& -6.793464  	& ($24.38\pm0.08,\,24.07\pm0.07,\,23.88\pm0.07,\,23.62\pm0.08$)\,\ \ \ \ \ \ \ \ \ \ \ \ \ \ \ \ \ \ \ \\
 A2       	 	& 2937882790 	& 35.46241   	& -6.793449  	& ($23.49\pm0.04,\,23.18\pm0.03,\,22.87\pm0.03,\,22.70\pm0.01$)\,\ \ \ \ \ \ \ \ \ \ \ \ \ \ \ \ \ \ \ \\
 \hline 
 
 \SystemBDES 	& 2925464779 	& 42.534995  	& 0.137950 		& ($24.83\pm0.18,\,23.04\pm0.18,\,21.74\pm0.02,\,21.10\pm0.01$)\ \ \ \ \ \ \ \ \ \ \ \ \ \ \ \ \ \ \ \\
 \arcA       	& 2925464554 	& 42.537078  	& 0.137222  	& ($24.11\pm0.10,\,23.72\pm0.06,\,23.38\pm0.07,\,22.79\pm0.04$)\ \ \ \ \ \ \ \ \ \ \ \ \ \ \ \ \ \ \ \\
 \arcB       	& 2925464667 	& 42.534792  	& 0.138905  	& ($24.26\pm0.11,\,24.06\pm0.08,\,23.81\pm0.10,\,23.37\pm0.08$)\ \ \ \ \ \ \ \ \ \ \ \ \ \ \ \ \ \ \ \\
\hline

 \SystemCDES  	& 2940549190 	& 52.369631  	& -28.327630 	& ($24.25\pm0.03,\,22.41\pm0.01,\,21.30\pm0.00,\,20.84\pm0.00,\,20.78\pm0.06$) \\
 \arcA 	  		& 2940548768 	& 52.367884  	& -28.327915 	& ($24.39\pm0.03,\,23.96\pm0.02,\,23.41\pm0.02,\,23.13\pm0.02,\,22.85\pm0.41$) \\
 \arcB       	& 2940548772 	& 52.368948  	& -28.325491 	& ($24.37\pm0.03,\,24.31\pm0.03,\,23.97\pm0.03,\,23.51\pm0.02$)\ \ \ \ \ \ \ \ \ \ \ \ \ \ \ \ \ \ \ \\ 
 \hline

 \SystemDDES 	& 3081734463 	& 52.737216  	& -52.470283 	& ($22.04\pm0.03,\,20.10\pm0.01,\,19.33\pm0.01,\,18.91\pm0.01,\,18.74\pm0.02$) \\
  \arcA     	& 3081733697 	& 52.735917  	& -52.471431 	& ($23.85\pm0.14,\,23.07\pm0.09,\,22.67\pm0.13,\,22.19\pm0.13,\,21.80\pm0.31$) \\
 \hline
  
 \SystemEDES  	& 2933341648 	& 71.553645  	& -51.442376 	& ($23.91\pm0.05,\,21.91\pm0.01,\,20.64\pm0.01,\,20.08\pm0.01,\,19.86\pm0.02$) \\
  \arcA 	  	& 2933341522 	& 71.557704  	& -51.442463	& ($23.24\pm0.02,\,22.64\pm0.02,\,22.33\pm0.03,\,22.11\pm0.04,\,22.28\pm0.15$) \\
  B1	  	  	& 2933341657 	& 71.552168  	& -51.440677 	& ($24.47\pm0.08,\,23.80\pm0.05,\,23.59\pm0.10,\,23.52\pm0.15,\,23.70\pm0.54$) \\
  B2      	 	& -		  	 	& 71.550762  	& -51.442894 	& ($24.15\pm0.07,\,23.36\pm0.05,\,23.11\pm0.04,\,22.78\pm0.04,\,23.10\pm0.23$)  \\
 \hline
 
 \SystemFDES  	& 2968302040 	& 354.029719 	& -53.876581 	& ($22.41\pm0.01,\,20.48\pm0.00,\,19.57\pm0.00,\,19.20\pm0.00,\,19.21\pm0.01$) \\
 \arcA 	  		& 2968302027 	& 354.02782  	& -53.876292 	& ($23.31\pm0.02,\,22.45\pm0.02,\,21.80\pm0.02,\,21.38\pm0.02,\,21.33\pm0.05$) \\
 \arcB       	& 2968301854 	& 354.032676 	& -53.878106 	& ($24.18\pm0.05,\,23.33\pm0.04,\,22.47\pm0.03,\,22.14\pm0.04,\,22.20\pm0.12$) \\
 [1ex]
 \hline \hline
\end{tabular} \\
\begin{flushleft}
{\bf Notes. } Positions and photometry of objects for each confirmed strong lensing system. \slitidname\ refers to the arc or lensed source image discussed in the text, in Fig.'s~\ref{fig:multipanel} and Fig.'s~\ref{fig:SystemA03}-\ref{fig:SystemF24}. The Object ID is the unique identification number in the DESDM database of coadd objects. All positions (R.A., Dec. in J2000) and magnitudes are drawn from the DESDM database, except in the case of source image B2 of \SystemEDES: for this object, we measured aperture magnitudes via the Graphical Astronomy and Image Analysis software tool \citep{gaiawebsite}. $Y$-band photometry is provided where available: supernova fields were only observed in $Y$-band for those fields that overlap with the wide-field survey, and arc B of \SystemCDES\ was too faint for detection in either DESDM and GAIA. 
\end{flushleft}
  \end{table*}
\normalsize
}

\def \tablelensingfeatures
{
\begin{table*}[!htb]
  \caption{Lensing Features \label{table:lensingfeatures}} 
  \centering 
\footnotesize
  \begin{tabular}{ ccccc } 
    \hline\hline             
   	 System ID  & Spectral    	&  Redshift  		& Einstein Radius   	& Enclosed Mass  		\\ 
	\slitidname & Features 		& 			        & $\thetae$ ($\arcsec$) & $\menclosed$ ($\msol$)\\ [0.5ex]                    
\hline  \hline    
\SystemADES  	& 				& $0.672\pm0.042$	& 						& 						\\
A1		 		& \lymanalpha  	& $\SystemAspeczA$  & $\SystemAradiusA$ 	& $\SystemAmencAtable$	\\ 
A2		 		& \lymanalpha  	& $\SystemAspeczB$  &  						& 						\\ 
\hline 
\SystemBDES  	& 				& $0.841\pm0.042$	& 						& 						\\
\arcA 		 	& \OII\        	& $\SystemBspeczA$  & $\SystemBradiusA$		& $\SystemBmencAtable$	\\ 
\hline 
\SystemCDES  	& 				& $0.655\pm0.033$	& 						& 						\\
\arcA 		 	& \OII\        	& $\SystemCspeczA$ 	& $\SystemCradiusA$ 	& $\SystemCmencAtable$ 	\\ 
\arcB 		 	& \OII\        	& $\SystemCspeczB$ 	& 						&  						\\ 
\hline 
\SystemDDES  	& 				& $0.463\pm0.046$	& 						& 						\\
\arcA 		 	& \OII\        	& $\SystemDspeczA$  & $\SystemDradiusA$ 	& $\SystemDmencAtable$ 	\\ 
\hline 
\SystemEDES  	& 				& $0.746\pm0.047$   & 						& 						\\
B1 		 		& \lymanalpha\ 	& $\SystemEspeczB$  & $\SystemEradiusB$ 	& $\SystemEmencBtable$ 	\\ 
B2 		 		& \lymanalpha\ 	& $\SystemEspeczC$ 	&  						&  						\\ 
\hline 
\SystemFDES  	& 				& $0.530\pm0.075$	& 						& \\
\arcA 		 	& \OII\        	& $\SystemFspeczA$ 	& $\SystemFradiusA$ 	& $\SystemFmencAtable$	\\ 
\arcB 		 	& \OII\        	& $\SystemFspeczB$ 	& $\SystemFradiusB$ 	& $\SystemFmencBtable$ 	\\ 
[1ex]
\hline \hline
 \end{tabular} \\ 
\begin{flushleft}
{\bf Notes. } Lensing features of confirmed systems. We show the DESDM object ID's for lenses, the source image IDs, names of spectral features, photometric redshifts of lenses $z_{\rm l}$,  spectroscopic redshifts of sources $z_{\rm s}$, an Einstein radius for each source image $\thetae$, and the resulting enclosed masses $\menclosed$. The principle spectral features are all emission lines. Redshifts for the lenses are the photometric redshifts drawn from the DESDM database: the uncertainties have been multiplied by 1.5 times the original estimate, according to the results of \cite{sanchez14} for estimating uncertainties in DES photometric redshift measurement codes. Redshifts for the sources are those measured from spectroscopic follow-up at Gemini South and Magellan. 
\end{flushleft}
\end{table*}
\normalsize
}
 
\def \tableredmapper
{
\begin{table*}[!htb]
  \caption{\redmapper\ Group and Cluster Properties \label{table:clusterproperties}} 
  \centering 
  \footnotesize
  \begin{tabular}{ c c c c } 
    \hline\hline             
    DES System ID 	&  Redshift      	& Richness     		& Mass  		 	    			\\ 
                  	& $z_{\rm RM}$      &	  		     	& $M_{\rm 500, RM}\ (10^{14} \msol)$     	 	  	\\ [0.5ex] 
\hline  \hline    
    \SystemADES     & $0.603\pm0.025$	& $4.6\pm1.2$		& $<.67$\footnote{The mass calibration method \citet{saro15} does not reach to low enough masses and richness to provide a complete probability distribution for the mass of this object. We instead cite an upper limit.}\\
    \SystemBDES 	& $0.841\pm0.013$   & $64.6\pm6.0$		& $3.72\pm0.71$\\
    \SystemCDES 	& $0.671\pm0.016$   & $57.7\pm3.4$ 		& $3.58\pm0.69$\\
    \SystemDDES 	& $0.435\pm0.013$   & $114.5\pm11.4$   	& $7.19\pm1.26$\\
    \SystemEDES 	& $0.730\pm0.030$ 	& $41.9\pm5.0$    	& $2.84\pm0.58$\\
    \SystemFDES 	& $0.521\pm0.010$   & $92.7\pm5.0$    	& $6.84\pm1.21$\\
     [1ex]
 \hline \hline
\end{tabular} \\
\begin{flushleft}
{\bf Notes. } Properties of \redmapper\ clusters of the confirmed strong lenses. The masses are measured with the methods of \citet{saro15}. 
\end{flushleft}
\end{table*}
\normalsize
}

\shorttitle{Strong Lenses in DES SV}
\shortauthors{Nord et al.}

\begin{document}

\title{Observation and Confirmation of Six Strong Lensing Systems in The Dark Energy Survey Science Verification Data}\thanks{This paper includes data gathered with the 6.5 meter Magellan Telescopes located at Las Campanas Observatory, Chile.}


\author{
B.~Nord\altaffilmark{1},
E.~Buckley-Geer\altaffilmark{1},
H.~Lin\altaffilmark{1},
H.~T.~Diehl\altaffilmark{1},
J.~Helsby\altaffilmark{2},
N.~Kuropatkin\altaffilmark{1},
A.~Amara\altaffilmark{3},
T.~Collett\altaffilmark{4},
S.~Allam\altaffilmark{1},
G.~Caminha\altaffilmark{5},
C.~De Bom\altaffilmark{5},
S.~Desai\altaffilmark{6,7},
H.~D\'umet-Montoya\altaffilmark{8},
 M.~Elidaiana da S. Pereira\altaffilmark{5},
D.~A.~Finley\altaffilmark{1},
B.~Flaugher\altaffilmark{1},
C.~Furlanetto\altaffilmark{9},
H.~Gaitsch\altaffilmark{1},
M.~Gill\altaffilmark{10},
K.~W.~Merritt\altaffilmark{1},
A.~More\altaffilmark{11},
D.~Tucker\altaffilmark{1},
E.~S.~Rykoff\altaffilmark{12,10},
E.~Rozo\altaffilmark{13},
F.~B.~Abdalla\altaffilmark{14,15},
A.~Agnello\altaffilmark{16},
M.~Auger\altaffilmark{17},
R.~J.~Brunner\altaffilmark{18,19},
M.~Carrasco~Kind\altaffilmark{18,19},
F.~J.~Castander\altaffilmark{20},
C.~E.~Cunha\altaffilmark{12},
L.~N.~da Costa\altaffilmark{21,22},
R.~Foley\altaffilmark{18,23},
D.~W.~Gerdes\altaffilmark{24},
K.~Glazebrook\altaffilmark{25},
J.~Gschwend\altaffilmark{21,22},
W.~Hartley\altaffilmark{3},
R.~Kessler\altaffilmark{2},
D.~Lagattuta\altaffilmark{26},
G.~Lewis\altaffilmark{27},
M.~A.~G.~Maia\altaffilmark{21,22},
M.~Makler\altaffilmark{5},
F.~Menanteau\altaffilmark{18,19},
A.~Niernberg\altaffilmark{28},
D.~Scolnic\altaffilmark{2},
J.~D.~Vieira\altaffilmark{18,23,19},
R.~Gramillano\altaffilmark{18},
T. M. C.~Abbott\altaffilmark{29},
M.~Banerji\altaffilmark{17,30},
A.~Benoit-L{\'e}vy\altaffilmark{31,14,32},
D.~Brooks\altaffilmark{14},
D.~L.~Burke\altaffilmark{12,10},
D.~Capozzi\altaffilmark{4},
A.~Carnero~Rosell\altaffilmark{21,22},
J.~Carretero\altaffilmark{20,33},
C.~B.~D'Andrea\altaffilmark{4,34},
J.~P.~Dietrich\altaffilmark{6,7},
P.~Doel\altaffilmark{14},
A.~E.~Evrard\altaffilmark{35,24},
J.~Frieman\altaffilmark{1,2},
E.~Gaztanaga\altaffilmark{20},
D.~Gruen\altaffilmark{36,37},
K.~Honscheid\altaffilmark{38,39},
D.~J.~James\altaffilmark{29},
K.~Kuehn\altaffilmark{40},
T.~S.~Li\altaffilmark{41},
M.~Lima\altaffilmark{42,21},
J.~L.~Marshall\altaffilmark{41},
P.~Martini\altaffilmark{38,43},
P.~Melchior\altaffilmark{38,44,39},
R.~Miquel\altaffilmark{45,33},
E.~Neilsen\altaffilmark{1},
R.~C.~Nichol\altaffilmark{4},
R.~Ogando\altaffilmark{21,22},
A.~A.~Plazas\altaffilmark{46},
A.~K.~Romer\altaffilmark{47},
M.~Sako\altaffilmark{48},
E.~Sanchez\altaffilmark{49},
V.~Scarpine\altaffilmark{1},
M.~Schubnell\altaffilmark{24},
I.~Sevilla-Noarbe\altaffilmark{49,18},
R.~C.~Smith\altaffilmark{29},
M.~Soares-Santos\altaffilmark{1},
F.~Sobreira\altaffilmark{1,21},
E.~Suchyta\altaffilmark{48},
M.~E.~C.~Swanson\altaffilmark{19},
G.~Tarle\altaffilmark{24},
J.~Thaler\altaffilmark{23},
A.~R.~Walker\altaffilmark{29},
W.~Wester\altaffilmark{1},
Y.~Zhang\altaffilmark{24}
\\ \vspace{0.2cm} (The DES Collaboration) \\
}

\affil{$^{1}$ Fermi National Accelerator Laboratory, P. O. Box 500, Batavia, IL 60510, USA}
\affil{$^{2}$ Kavli Institute for Cosmological Physics, University of Chicago, Chicago, IL 60637, USA}
\affil{$^{3}$ Department of Physics, ETH Zurich, Wolfgang-Pauli-Strasse 16, CH-8093 Zurich, Switzerland}
\affil{$^{4}$ Institute of Cosmology \& Gravitation, University of Portsmouth, Portsmouth, PO1 3FX, UK}
\affil{$^{5}$ ICRA, Centro Brasileiro de Pesquisas F\'isicas, Rua Dr. Xavier Sigaud 150, CEP 22290-180, Rio de Janeiro, RJ, Brazil}
\affil{$^{6}$ Excellence Cluster Universe, Boltzmannstr.\ 2, 85748 Garching, Germany}
\affil{$^{7}$ Faculty of Physics, Ludwig-Maximilians University, Scheinerstr. 1, 81679 Munich, Germany}
\affil{$^{8}$ Universidade Federal do Rio de Janeiro - Campus Maca\'e, Rua Alo\'isio Gomes da Silva, 50 - Granja dos Cavaleiros, Cep: 27930-560, Maca\'e, RJ, Brazil}
\affil{$^{9}$ University of Nottingham, School of Physics and Astronomy, Nottingham NG7 2RD, UK}
\affil{$^{10}$ SLAC National Accelerator Laboratory, Menlo Park, CA 94025, USA}
\affil{$^{11}$ Kavli IPMU (WPI), UTIAS, The University of Tokyo, Kashiwa, Chiba 277-8583, Japan}
\affil{$^{12}$ Kavli Institute for Particle Astrophysics \& Cosmology, P. O. Box 2450, Stanford University, Stanford, CA 94305, USA}
\affil{$^{13}$ Department of Physics, University of Arizona, Tucson, AZ 85721, USA}
\affil{$^{14}$ Department of Physics \& Astronomy, University College London, Gower Street, London, WC1E 6BT, UK}
\affil{$^{15}$ Department of Physics and Electronics, Rhodes University, PO Box 94, Grahamstown, 6140, South Africa}
\affil{$^{16}$ Department of Physics and Astronomy, PAB, 430 Portola Plaza, Box 951547, Los Angeles, CA 90095-1547, USA}
\affil{$^{17}$ Institute of Astronomy, University of Cambridge, Madingley Road, Cambridge CB3 0HA, UK}
\affil{$^{18}$ Department of Astronomy, University of Illinois, 1002 W. Green Street, Urbana, IL 61801, USA}
\affil{$^{19}$ National Center for Supercomputing Applications, 1205 West Clark St., Urbana, IL 61801, USA}
\affil{$^{20}$ Institut de Ci\`encies de l'Espai, IEEC-CSIC, Campus UAB, Carrer de Can Magrans, s/n,  08193 Bellaterra, Barcelona, Spain}
\affil{$^{21}$ Laborat\'orio Interinstitucional de e-Astronomia - LIneA, Rua Gal. Jos\'e Cristino 77, Rio de Janeiro, RJ - 20921-400, Brazil}
\affil{$^{22}$ Observat\'orio Nacional, Rua Gal. Jos\'e Cristino 77, Rio de Janeiro, RJ - 20921-400, Brazil}
\affil{$^{23}$ Department of Physics, University of Illinois, 1110 W. Green St., Urbana, IL 61801, USA}
\affil{$^{24}$ Department of Physics, University of Michigan, Ann Arbor, MI 48109, USA}
\affil{$^{25}$ Centre for Astrophysics \& Supercomputing, Swinburne University of Technology, Victoria 3122, Australia}
\affil{$^{26}$ Centre de Recherche Astrophysique de Lyon, Universit\'e de Lyon, Universit\'e Lyon 1, CNRS, Observatoire de Lyon; 9 avenue Charles Andr\'e, 69561 Saint-Genis Laval Cedex, France}
\affil{$^{27}$ Sydney Institute for Astronomy, School of Physics A28, The University of Sydney, NSW 2006, Australia}
\affil{$^{28}$ Center for Cosmology and Astro-Particle Physics, The Ohio State University, Columbus OH 43210, USA}
\affil{$^{29}$ Cerro Tololo Inter-American Observatory, National Optical Astronomy Observatory, Casilla 603, La Serena, Chile}
\affil{$^{30}$ Kavli Institute for Cosmology, University of Cambridge, Madingley Road, Cambridge CB3 0HA, UK}
\affil{$^{31}$ CNRS, UMR 7095, Institut d'Astrophysique de Paris, F-75014, Paris, France}
\affil{$^{32}$ Sorbonne Universit\'es, UPMC Univ Paris 06, UMR 7095, Institut d'Astrophysique de Paris, F-75014, Paris, France}
\affil{$^{33}$ Institut de F\'{\i}sica d'Altes Energies (IFAE), The Barcelona Institute of Science and Technology, Campus UAB, 08193 Bellaterra (Barcelona) Spain}
\affil{$^{34}$ School of Physics and Astronomy, University of Southampton,  Southampton, SO17 1BJ, UK}
\affil{$^{35}$ Department of Astronomy, University of Michigan, Ann Arbor, MI 48109, USA}
\affil{$^{36}$ Max Planck Institute for Extraterrestrial Physics, Giessenbachstrasse, 85748 Garching, Germany}
\affil{$^{37}$ Universit\"ats-Sternwarte, Fakult\"at f\"ur Physik, Ludwig-Maximilians Universit\"at M\"unchen, Scheinerstr. 1, 81679 M\"unchen, Germany}
\affil{$^{38}$ Center for Cosmology and Astro-Particle Physics, The Ohio State University, Columbus, OH 43210, USA}
\affil{$^{39}$ Department of Physics, The Ohio State University, Columbus, OH 43210, USA}
\affil{$^{40}$ Australian Astronomical Observatory, North Ryde, NSW 2113, Australia}
\affil{$^{41}$ George P. and Cynthia Woods Mitchell Institute for Fundamental Physics and Astronomy, and Department of Physics and Astronomy, Texas A\&M University, College Station, TX 77843,  USA}
\affil{$^{42}$ Departamento de F\'{\i}sica Matem\'atica,  Instituto de F\'{\i}sica, Universidade de S\~ao Paulo,  CP 66318, CEP 05314-970, S\~ao Paulo, SP,  Brazil}
\affil{$^{43}$ Department of Astronomy, The Ohio State University, Columbus, OH 43210, USA}
\affil{$^{44}$ Department of Astrophysical Sciences, Princeton University, Peyton Hall, Princeton, NJ 08544, USA}
\affil{$^{45}$ Instituci\'o Catalana de Recerca i Estudis Avan\c{c}ats, E-08010 Barcelona, Spain}
\affil{$^{46}$ Jet Propulsion Laboratory, California Institute of Technology, 4800 Oak Grove Dr., Pasadena, CA 91109, USA}
\affil{$^{47}$ Department of Physics and Astronomy, Pevensey Building, University of Sussex, Brighton, BN1 9QH, UK}
\affil{$^{48}$ Department of Physics and Astronomy, University of Pennsylvania, Philadelphia, PA 19104, USA}
\affil{$^{49}$ Centro de Investigaciones Energ\'eticas, Medioambientales y Tecnol\'ogicas (CIEMAT), Madrid, Spain}

\altaffiltext{1}{B.N. email: nord@fnal.gov}

\begin{abstract}
We report the observation and confirmation of the first group- and cluster-scale strong gravitational lensing systems found in Dark Energy Survey (DES) data. Through visual inspection of data from the Science Verification (SV) season, we identified 53 candidate systems. We then obtained spectroscopic follow-up of 21 candidates using the Gemini Multi-Object Spectrograph (GMOS) at the Gemini South telescope and the Inamori-Magellan Areal Camera and Spectrograph (IMACS) at the Magellan/Baade telescope. With this follow-up, we confirmed six candidates as gravitational lenses:  Three of the systems are newly discovered, and the remaining three were previously known. Of the 21 observed candidates, the remaining 15 were either not detected in spectroscopic observations, were observed and did not exhibit continuum emission (or spectral features), or were ruled out as lensing systems. The confirmed sample consists of one group-scale and five galaxy cluster-scale lenses. The lensed sources range in redshift $z\sim0.80-3.2$, and in $i$-band surface brightness $i_{\rm SB} \sim23-25\ {\rm mag}/\sqarcsec$ ($2\arcsec$ aperture). For each of the six systems, we estimate the Einstein radius $\thetae$ and the enclosed mass $\menclosed$, which have ranges $\thetae \sim 5.0 - 8.6\arcsec$ and $\menclosed \sim 7.5 \times 10^{12} -  6.4\times 10^{13}\, \msol$, respectively. 
\end{abstract}


\section{Introduction}\label{sec:introduction}

\figuremultipanel

Strong gravitational lensing of galaxies and quasars provides opportunities to study cosmology, dark matter, dark energy, galactic structure and galaxy evolution \citep{treu10}. Strong lensing also provides a sample of galaxies---the lenses themselves---that are selected based on total mass, rather than luminosity or surface brightness  \citep[e.g.,][]{reblinsky99}: the distortion of images by the lens  depends only on the total lens mass and its spatial distribution\footnote{This is not the case for strong lensing systems with multiple lens planes spread out across the distance between the lens and source, but these cases are relatively rare.}.

The majority of strong lensing systems discovered in the last decade were found with a variety of techniques, mostly through dedicated investigations of existing data or through new dedicated surveys. Data from the Sloan Digital Sky Survey \citep[SDSS;][]{york00}, for example, have been used by a number of groups to select lens candidates. The Optimal Line-of-Sight-Lens Survey \citep[OLS;][]{willis06} discovered five galaxy-scale lenses. Investigating  a sample of rich galaxy clusters, \citet{hennawi08} found 37 lenses of varying size and morphology. The Sloan Bright Arc Survey \citep[SBAS; ][]{allam07, kubik07, diehl09,lin09, kubo09, kubo10, wiesner12} discovered 28 galaxy- and group-scale systems.   The Sloan Lens ACS Survey group \citep[SLACS;][]{bolton08} identified 137 lensing candidates, using both the spectroscopic and imaging catalogs.  The CAmbridge Sloan Survey Of Wide ARcs in the skY \citep[CASSOWARY;][]{belokurov09} searched for wide-separation, galaxy-scale systems, uncovering 45 systems to date. Other searches, like the Strong Lensing Legacy Survey \citep[SL2S; e.g.,][]{cabanac07,more12}, have yielded 127 lens candidates---spanning group to cluster scales---from $\sim150$ \sqdeg\ of sky area in the Canada-France-Hawaii Telescope Legacy Survey (CFHTLS).  Searches in the COSMOS field \citep{faure08,jackson08} have yielded 70 lens candidates. 

Surveys in other bands have also proven successful. The Cosmic Lens All-Sky Survey\footnote{\url{http://www.aoc.nrao.edu/~smyers/class.html}} \citep[CLASS; e.g.,][]{myers03, browne03, york05} confirmed nearly 25 radio-loud galaxy-scale lensing systems. Several of them have multiply-imaged sources, including the four-image system CLASS B2045+265 \citep{fassnacht99}, and the two-component wide-separation (4.56$\arcsec$) system CLASS B2108+213 \citep{mckean05}. 

There is a history of targeting massive galaxy clusters to use as cosmic lenses for the study of high-redshift star-forming galaxies in the infrared. \citep[e.g.,][]{smail97, swinbank10, egami10}. Recently, wide-field millimeter and submillimeter surveys have enabled the discovery of new populations of strong lensing systems based on a simple flux-density selection \citep{blain96,negrello07} with near-100\% completeness \cite{vieira10,negrello10,bussmann13}. These wide-field surveys, coupled with the recently commissioned Atacama Large Millimeter Array (ALMA), have opened a new window to star-formation at high redshifts \citep[e.g.,][]{vieira13,alma15}.

These investigations---from optical to millimeter wavebands---produced on the order of 1000 lensing candidates or confirmed lenses, spanning a range of arc sizes---from small arcs ($\thetae \sim 3\arcsec$) produced by single-galaxy lenses to giant arcs typically found in clusters ($\thetae \sim 10\arcsec$). 

The Dark Energy Survey \citep[DES\footnote{\url{www.darkenergysurvey.org}};][]{diehl14,flaugher15}---a new, deep sky survey covering 5000 \sqdeg\ of the southern Galactic Cap in five optical filters ($grizY$)---is suitable for a systematic census of gravitational lenses. The wide-field survey will have a depth of $i\sim24$ mag at $10\sigma$ detection threshold, and the 30\sqdegadjective\ supernova survey is spread over 10 fields --- eight shallow and two deep. The primary instrument for DES is the newly assembled Dark Energy Camera (DECam), a wide-field (3 \sqdeg) CCD mosaic camera located on the Blanco 4m telescope at the Cerro Tololo Inter-American Observatory in the Chilean Andes. The DES footprint is best observed between August and mid-February. DES observes for $\sim105$ nights per year, and has completed two of the five years of planned observations. The survey will produce the widest, deepest, and most uniform contiguous map in optical wavelengths of the southern sky, much of which has not been systematically surveyed---and not at such depth---until now. These features make DES suitable for finding many strong lensing systems that span large ranges in redshift and Einstein radii.  DES expects to uncover a large and diverse sample of strong gravitational lenses, which will provide for a rich science program. 

\tablelensobjects

One of the main objectives of the strong lensing science program in DES is to derive constraints on dark energy. One of the two key components will be exploiting galaxy- and cluster-scale lenses that have multiple background sources at different redshifts. When there exist two lensed sources, one source will lie between the lens and the other, more distant source: these sources produce distinct Einstein radii. The ratio of the Einstein radii (and thus, a ratio of angular diameter distances) is independent of the Hubble constant, and only weakly dependent on the lens mass distribution.  Provided the lensing mass can be well modeled, the ratio of distance ratios can help to constrain dark energy models \citep{gavazzi08, jullo10, collett12, collett14}. 

The second key component is to use strongly lensed quasars to measure cosmological parameters (e.g., $ H_0$) by utilizing the time delays between multiple images \citep{refsdal64,suyu13} of time-varying sources.   We predict that the wide-field survey should contain about 120 lensed quasars brighter than $i = 21$ mag \citep{oguri10}, of which $\sim20$ would be high-information content quadruple-image configurations. Through the STRong-lensing Insights into Dark Energy Survey (STRIDES) program\footnote{\url{http://strides.physics.ucsb.edu/}}, DES has discovered and confirmed two lensed quasars  so far at $z\sim1.6$ and $\sim2.4$ \citep{agnello15}. In addition, 30~\sqdeg\ of deep imaging in the supernova fields, which have a high cadence, provides an opportunity to find lensed supernovae.

In addition to constraining cosmology, we will use a substantial cluster-scale lens sample to study dark matter mass profiles. Also, we expect to discover a large sample of sources at varying redshifts that will be valuable for studies of galaxy evolution.

In this paper, we report on the search for strong lens systems from the DES SV data. We performed a visual inspection of the $\sim250$\sqdegadjective\ SV area, discovering a large number of candidates, which we then culled to 53 high-ranking candidates. We selected 24 for spectroscopic follow-up and were able to observe 21 of them.  Of the sample we followed up, we focus on six systems that have compelling evidence (color, morphology and spectroscopy) of strong gravitational lensing: three of these are newly identified and three are already known from other surveys. We have spectroscopically confirmed these six systems (Fig.~\ref{fig:multipanel}) as strong gravitational lensing systems.

Several of the candidate systems were selected for the possibility of housing sources at multiple redshifts, key for our dark energy studies. While we do not endeavor in this work to perform detailed mass modeling of cluster-scale lenses with DES data, future higher-resolution imaging follow-up may allow for detailed modeling of cluster mass distributions.

The paper is organized as follows. In \S\ref{sec:svdata} we describe the DES SV season data. In \S\ref{sec:search}, we present the lens search followed by the spectroscopic follow-up in \S\ref{sec:follow-up}. We present the six confirmed systems and their properties in \S\ref{sec:sample}, discuss results in \S\ref{sec:discussion}, and conclude in \S\ref{sec:summary}. All magnitudes are in the AB system. We assume a flat, Planck $\Lambda$CDM cosmology: $\Omega_{\rm M} =0.308$,  $\Omega_{\Lambda} = 0.692$, and $H_0=  67.8\, \kms\,{\rm Mpc}^{-1}$ \citep{planckcosmo15}.

\section{DES Science Verification (SV) Data}\label{sec:svdata}

The SV observing season took place from November 1, 2012 to February 23, 2013, immediately after the commissioning of the Dark Energy Camera \citep{flaugher15}. The survey started in August 2013 and has completed two seasons beyond SV. The survey and operations for SV and Year 1 are described in \cite{diehl14}.

In total, SV observations cover $\sim 250$ \sqdeg\ in five optical filter bands ($grizY$), nearly reaching the expected full five-year depth of DES: below, we report median depth estimates of these data at $10\sigma$ galaxy limiting magnitude. The wide-field survey consists of the SPT-E and SPT-W fields that are located, respectively, in the east and west regions of the South Pole Telescope (SPT) survey footprint \citep{carlstrom11}: the $g$-band depth extends to $\sim 24.5$ mag. The $30$\sqdegadjective\ supernova survey takes place in eight shallow fields, and two deep fields, extending to $g$-band depths, $25.0$ mag and $25.4$ mag, respectively.

The DES data management (DESDM) team was responsible for the reduction of the SV images. The basic reductions and coadd source detection were performed by DESDM \citep{desai12, gruendl15}: the software for reduction of SV exposures is similar to that used in the Y1A1 release\footnote{\url{http://data.darkenergysurvey.org/aux/releasenotes/DESDMrelease.html}}. The detrended, single-epoch images were calibrated, background-subtracted, and combined into ‘coadd tiles’. Each coadd tile has dimensions $0.72\ {\rm deg.\ }\ \times 0.72\ {\rm deg.\ }$, so defined as to cover the entire DES footprint. A catalog of objects was extracted from the coadded images using Source Extractor \citep[SExtractor;][]{bertin96, bertin11}. Unless otherwise stated, photometric and position data of all objects in this work originate in this object catalog (see Table~\ref{table:lensingobjects}).

\tableobservationlog

\section{Lens Search}\label{sec:search}

The lens search, carried out by visually inspecting images, relied on morphology and color to identify lens candidates. We searched for arc- and ring-like features, and multiply imaged sources, in association with red lensing galaxies or galaxy groups. Among the more typical of search criteria is the existence of blue source images in association with red galaxies, but some systems with red source images were also identified. 

Although modern algorithms for automated arc- and ring-finders have improved in recent years, they still typically require manual intervention and/or subsequent visual inspection and verification \citep{lenzen04,horesh05,seidel07, more12, joseph14}. We performed a visual search without automated finders, because the SV sky area is relatively small and visual inspection is likely to recover the highest-ranking candidates that would be found with an automated finder. Given these search criteria, as well as the pixel scale of DECam and the seeing of DES data, there is a bias towards larger Einstein radii, and thus higher mass (depending on the source redshift(s)), which are easier to identify through visual inspection. 

We performed two types of visual scans.  We performed a \opensearch\ search of all $\sim250$ \sqdeg---not focusing on any specific target regions or objects---as well as a targeted scan of previously identified galaxy clusters. The visual scan is performed on false-color PNG images, which are made by combining $g,r,i$ coadd tiles (defined in \S\ref{sec:svdata}) into color images. 

The \opensearch\ search was carried out by about 20 inspectors, who systematically examined all the tiles. The open-area search yielded a sample of $\sim 1000$ objects. Among the results of this search is the Cosmic Horseshoe \citep{belokurov07}, which was then removed from our sample of follow-up candidates. 

We also carried out two targeted searches. The first one was carried out around the 67 known clusters from the SPT SZ survey \citep{bleem15}. Many of these clusters already have optical imaging follow-up and known lensing features. We identified nine candidates from this search. The second targeted search examined 374 galaxy clusters of richness $ > 50$,  which themselves were found using the red-sequence Matched-Filter Probabilistic Percolation cluster-finder algorithm \citep[\redmapper;][]{rykoff14}. This search yielded 39 candidates. The targeted search recovered known, confirmed lensing systems, such as BCS J2352-5452 \citep{buckleygeer11}, RXC J2248.7-4431 \citep{bohringer04} and  ACT-CL J0102-4915 \citep[El Gordo;][]{menanteau12}, which we then removed from the sample of candidates for spectroscopic follow-up. 

The combined \opensearch\ and targeted search samples were merged, and assessed by a team of experienced inspectors. Most of these objects were not convincing as strong lens system candidates or were not bright enough to expect successful spectroscopic follow-up. Each of the systems was assigned a rank, 1 to 3---with 1 signifying least likely to be a lens system, and 3 the most likely---on the basis of  morphology, color, and brightness of sources. This ranking resulted in 53 promising (`rank 3') candidates.

To characterize the selection function of our sample, we compare to a sample of 87 high-quality candidates and confirmed lensing systems found in CFHTLS data. It includes a sample of 33 confirmed lenses from the \textsc{RingFinder} \citep{gavazzi14}, the 26 most promising candidates from \textsc{ArcFinder} as implemented in SL2S ARCS \citep[SARCS;][]{more12}, and 28 candidates from \textsc{SpaceWarps} \citep{more16}, giving a total of 87 objects; the Einstein radii have a range, $1\arcsec < \thetae < 18\arcsec$. In the \textsc{ArcFinder} search, 12 candidates have giant arcs, which canonically have length-to-width ratio $\geq 8$. CFHTLS has an effective search area of $\sim150$ \sqdeg, and depths of $g\sim 25.5$ mag and $\sim 26.5$ mag in wide and deep fields, respectively. 

We estimate the lens-finding selection function of DES by counting the sources from the high-quality CFHTLS sample that survive a cut on the DES magnitude limit, which is taken to be $24.5$ mag in $g$-band for both the SV and full wide-field data. The source magnitudes of the samples listed above were provided by the author (Anupreeta More, personal communication, November 11, 2015), and the calculation of the source magnitudes is discussed in Section 5.3 \cite{more16}. After performing the magnitude cut, we linearly scaled the resulting counts by the ratio of the survey areas. From this comparison, we estimate that the full DES Wide-field survey will contain 2100 promising candidates. Scaling the area again for the DES SV area, we expect 105 promising candidates, which is nearly twice the amount found in the search discussed in this work.

A few factors likely contributed to the difference between the number of lens system candidates expected and discovered in DES SV data. The non-targeted search was performed by scanning large images, rather than cut-outs around particular objects, as was done with the targeted search around clusters. In addition, some regions were only scanned once during the non-targeted search. Rigorous characterization of a strong lens candidate selection function requires the construction of a framework to measure each scanner's efficiency, completeness, and purity. Such a program was not implemented for the SV searches.

\section{Spectroscopic Follow-up}\label{sec:follow-up}

We obtained spectroscopic follow-up observations with the Gemini Multi-Object Spectrograph \citep[GMOS;][]{2004PASP..116..425H} on the Gemini South Telescope, as part of the Gemini Large and Long Program GS-2014B-LP-5\footnote{\url{http://www.gemini.edu/?q=node/12238\#Buckley}}, and with the IMACS multi-object spectrograph \citep{dressler11} on the Magellan/Baade Telescope. 

Of the 53 high-ranking candidates found during the lens search, we selected the $24$ candidates most suitable for spectroscopic follow-up. For the purposes of planning the follow-up, the $24$ candidates were ranked (1-3) by their surface brightness, which is calculated using a $2\arcsec$-diameter aperture in the $i$-band: 
\begin{itemize}
\item R1: $i_{\rm SB} < 23\, {\rm mag}/\sqarcsec$ (six candidates)
\item R2: $23 < i_{\rm SB} < 24\, {\rm mag}/\sqarcsec$ (14 candidates)
\item R3: $i_{\rm SB} >24\, {\rm mag}/\sqarcsec$ (four candidates)
\end{itemize}
These rankings are not related to the quality rankings of candidates in \S\ref{sec:search}. Details of the observing parameters and conditions for each of our confirmed lensing systems are shown in Table~\ref{table:observationlog}. We acquired spectroscopic follow-up for 21 of the 24 candidate systems: 17 systems were observed at Gemini and five at Magellan. One of the systems, \SystemCDES, was observed at both Gemini and Magellan: only B600 data were acquired at Gemini, which presented no spectral features, so we observed at longer wavelengths at Magellan. 

We searched the Master Lens Database\footnote{\url{masterlens.astro.utah.edu}} and the Gemini Science Archive\footnote{\url{http://www.cadc-ccda.hia-iha.nrc-cnrc.gc.ca/en/gsa/index.html}} to ascertain if any of our candidates had been previously discovered or confirmed. The Master Lens Database (updated in April, 2014) contains candidate SA39 \citep{more12}, discovered by the SL2S in CFHTLS data;  we have obtained the first spectroscopic confirmation that SA39 is a strong lensing system. This is \SystemADES\ in our sample, and we present its confirmation in \S\ref{sec:sample:systemA03}.

\subsection{Gemini/GMOS}\label{sec:followup:gemini}

We observed some of the lensing candidates at Gemini South using the multi-object mode on GMOS. The mask for each field was centered on the candidate lens of each system, and slits were placed on the candidate source images. In some cases, it was necessary to shift the field center to locate a suitable guide star. Accommodating as many source targets as possible sometimes required rotation of the slit mask (i.e., rotation of the position angle of each system). We placed slits on as many of the lensing features as possible, and the remaining $\sim50$ slits were placed on galaxy targets for photometric-redshift calibration of DES data. The slits were $1\arcsec$ in width and of varying length in order to accommodate both the object and an amount of sky sufficient to perform reliable background subtraction. In some cases, we tilted the slits to maximize the captured flux. We defined three sets of exposure times/grating combinations, which depend on the surface brightness classes defined above:

\begin{itemize}
\item R1: 1-hour integration with R150/GG455 grating/filter. 
\item R2: 1-hour integration with B600 grating and 3.7-hour integration with R150/GG455 grating/filter.
\item R3: 1-hour integration with B600 grating.
\end{itemize}

We use the R150 grating in conjunction with the GG455 filter in order to obtain spectra with wavelength coverage $\sim4500-10000$\AA. For the sources that we expect to be late-type emission line galaxies, this would allow us, in most cases, to detect \OII\ to $z\sim 1.7$, $\Hbeta$ to $z\sim 1.0$ and \lymanalpha\ in the range $z\sim 2.7 - 7.2$. We use the B600 grating to obtain spectral coverage of $3250-6250$\AA, which would allow us to detect sources with $z > 2.0$ that emit \lymanalpha.

A 1-hour observing sequence consisted of a pair of 900-second exposures, followed by a flat field taken with a Quartz-Halogen lamp and a calibration spectrum taken with a CuAr arc lamp. We then dithered to a different central wavelength to cover the gap between the CCDs and took a CuAr spectrum, followed by the flat-field exposure and then two more 900-second exposures. The 3.7-hour integration used the same sequence repeated 16 times, but with 840-second exposure times. Dividing the integration time into multiple exposures facilitates the removal of cosmic rays. The data were binned $2\times2$, giving effective dispersions of $0.1$ and $0.386$ nm/pixel for the B600 and R150 gratings, respectively. The low-surface brightness systems (i.e., rank R3) underwent only one hour of integration with the B600 grating. The goal of this work is to confirm and further analyze a set of strong lensing systems: to conserve telescope time and maximize the number of lensing system confirmations, we elected to not pursue longer integrations for a given candidate system, even if a clear spectroscopic signal did not appear. In addition, the photo-z calibration targets require observations with consistent depth, and thus integration time, across the observation fields.

We used the Gemini IRAF package v2.16\footnote{\url{http://www.gemini.edu/sciops/data-and-results/processing-software}} to reduce all exposures.  In each system, for each wavelength dither, we first process the flat field using the {\tt gsflat} task (this includes subtraction of the bias). Each science exposure in a single dither is then reduced with {\tt gsreduce} (using the previously processed flat fields), and then the two exposures are combined with {\tt gemcombine}. Wavelength calibration and transformation are performed on each dither (using the {\tt gswavelength} and {\tt gstransform} tasks) before the pairs of dithers are coadded on a common wavelength scale to eliminate CCD chip gaps. We perform sky subtraction and 1D spectral extraction using {\tt gsextract}, which employs the {\tt apall} task. Feature identification and redshift estimation are performed using the {\tt emsao} task within the {\tt rvsao} IRAF package \citep{kurtz98}. We modified some of the Gemini IRAF tasks to provide more flexibility in the data reduction. In particular, we modified {\tt gswavelength} and {\tt gsextract} to allow one to perform 1D and 2D fits on the calibration data and sky subtraction on the science data, respectively, for individual extensions.

Of the 21 observed candidate systems, 17 were observed at Gemini: 12 systems were observed in Priority Visiting observer mode from October 19-24, 2014; and five systems were observed in queue mode. For nine systems, we obtained all the planned exposures, and for eight systems, we obtained a subset of the planned exposures. The only planned 1-hour integrations acquired with the R150 grating (rank R1) were executed during the Priority Visiting Observer mode. The eight systems for which we did not obtain all planned exposures are missing the planned 3.7-hour R150 observations: in queue mode, these observations had lower Gemini observing priority. All planned 1-hour B600 observations were completed.

\subsection{Magellan/IMACS}
We also obtained spectra at Magellan with IMACS. The mask was again centered on the candidate lensing galaxy of each system. As with GMOS, we used $1\arcsec$-width slits of varying length. We used the f/2 camera, and the 200-$\ell$/mm grism, because it has the best response in the redder wavelengths, and we used the WB5650-9200 filter to set the wavelength range to 5650-9200\AA.  We took three 1200-second exposures for each mask to facilitate cosmic ray removal. Two of the systems observed also had a 1-hour integration using the B600 grating using Gemini's GMOS. A ranking system was not used to select which systems to prioritize for observation.

To process the data we used the COSMOS data reduction package\footnote{\url{http://code.obs.carnegiescience.edu/cosmos}}. This package performs wavelength calibration, followed by subtraction of bias, and division by flats, using the programs, {\tt Sflats} and {\tt biasflat}. The sky is then subtracted using the program, {\tt subsky}. Finally extraction of individual one- or two-dimensional spectral exposures is performed using {\tt extract}, or {\tt extract-2dspec}. Multiple sets of extracted spectra are then combined using {\tt sumspec}. Feature identification and redshift estimation are  performed using the EMSAO IRAF task.

Of the 21 observed candidate systems, five were observed on December 20-21, 2014 as part of a larger program to obtain spectroscopic redshifts of 1) supernova hosts and 2) galaxies for photometric redshift calibration of DES data. Note that one system, \SystemCDES, was observed at both Magellan and Gemini.

\section{Confirmed Lens Sample}\label{sec:sample}
Of the 24 systems for which we sought follow-up, there was sufficient time to observe 21 systems. Of those, we confirmed a total of six strong lensing systems. There are 15 systems that were observed, but were not confirmed as lenses: nine have no discernible continuum emission (nor spectral features); four have continuum emission but no discernible features; and two were ruled out as lenses.

A number of the systems that we failed to confirm exhibit clear morphological and photometric lensing features. However, they would require significantly longer integration times or are located in the redshift desert---outside the range of the optical observations in our observing program. This highlights the challenge of obtaining spectra for the very faint objects found in deep optical surveys: a great deal of integration time, in tandem with IR spectrographs, may be required to confirm the large lens samples predicted in current and future large surveys. The candidates that were ruled out as lensing systems indeed exhibit suggestive morphological features, but the candidate lensed sources were determined to be foreground galaxies.

A multi-panel plot of the six confirmed systems is shown in Fig.~\ref{fig:multipanel}, and Table~\ref{table:lensingobjects} lists the positions and photometry of the candidate lens and source(s) of each lensing system observed. The sample is comprised of one group-scale lens and five cluster-scale lenses. The group-scale lens \SystemADES\ is known from the SARCS sample by SL2S \citep{more12}, but had not been spectroscopically confirmed by that group. It was found during the non-targeted search and later found to be of low richness ($<$5) in the \redmapper\ catalog. The remaining four (\SystemBDES, \SystemCDES, \SystemEDES\ and \SystemFDES) were  also identified in the \redmapper\ cluster-finder sample.  \SystemDDES\ \citep[SPT-CL J0330-5228/ACT-CL J0330-5227;][]{bleem15} and \SystemFDES\ \citep[SCSO J233607-535235;][]{menanteau10} are already known candidate lensing systems from SPT/ACT and SCS surveys, respectively, but there were previously no spectroscopic observations of their lensing features.  \SystemADES\ and \SystemBDES\ were found in the slightly deeper DES supernova fields during the non-targeted searches. 


In this section, we comment on each system, its spectral features of interest and the derived redshifts. The reduced 1D spectra, alongside cut-outs of the fields centered on the central lensing object showing the slit positions, targets and Einstein radii, are shown for each system in Fig.'s~\ref{fig:SystemA03}-\ref{fig:SystemF24}. The limitations of the ground-based images and the complexity of many of the systems preclude detailed mass modeling. Therefore, for each system, we provide estimates of the Einstein radius (or radii if there is more than one source) by manually fitting a circle that passes through one or more spectroscopically confirmed features. The center is chosen to be the center of curvature of the candidate arc image(s). We then estimate the enclosed mass $\menclosed$ of the lensing system, assuming a singular isothermal sphere (SIS) mass profile \citep{narayan96}:
\begin{equation}
\menclosed = \frac{c^2}{4G} \thetae^2 \left(\frac{D_{\rm L} D_{\rm S}}{D_{\rm L S}}\right), \label{eqn:menclosed}
\end{equation}
where $c$ is the speed of light, $G$ is Newton's gravitational constant, and $\thetae$ is the Einstein radius. $D_{\rm L}$, $D_{\rm S}$, and $D_{\rm LS}$, are angular diameter distances to the lens, to the source, and between the lens and source, respectively; they depend only on the redshifts to these objects and on the cosmological parameters. The spectral features, photometric redshifts of lenses, spectroscopic redshifts of sources, Einstein radii and enclosed masses are summarized in Table~\ref{table:lensingfeatures}. 

All of the confirmed systems are associated with \redmapper\ groups or clusters. The cluster masses were measured with the procedure of \cite{saro15}, which uses \redmapper\ and SPT clusters to calibrate richness and mass. The spherical overdensity mass $M_{\Delta}$ is that enclosed within a radial boundary demarcated by a density contrast $\Delta$ with respect to the critical density. We summarize the redshifts, richnesses and masses of these clusters in Table~\ref{table:clusterproperties}.

We estimate frequentist uncertainties for the enclosed mass, which include uncertainties propagated from redshifts (via the angular diameter distances) and Einstein radii. The photometric redshift uncertainties have been multiplied by a factor of $1.5$, according to the prescription of \citet{sanchez14} for photometric redshift measurements in DES. The redshift uncertainties are the result of a sum in quadrature of the wavelength calibration uncertainty and the redshift determination uncertainty from the IRAF function, EMSAO. The dominant contributor to uncertainties varies from system to system: for some, the photometric redshift uncertainties dominate and for others, it is the Einstein radius. As a test, we reduce the uncertainties on the Einstein radii artificially to $0.3\arcsec$: most of the mass estimates still have $> 50\%$ uncertainties.

\subsection{\SystemADES}\label{sec:sample:systemA03}
\SystemADES\ was originally discovered in CFHTLS \citep[Object SA39;][]{more12} and encountered during our non-targeted search. It was also serendipitously found to be a group-scale lens in the \redmapper\ catalog, with richness, $4.64\pm1.15$ and redshift, $0.603\pm0.025$. The central red galaxy of the lens has a DESDM photometric redshift of $\zlens=0.672\pm 0.042$.  There are two prominent blue arcs, A1 and A2, shown in Fig.~\ref{fig:multipanel}a, to the southeast and northeast of the lensing galaxy, respectively. These arcs are also shown in the top and middle panels on the right of Fig.~\ref{fig:SystemA03}. In the follow-up GMOS spectroscopy of both arcs, we identify strong  emission lines (Fig.~\ref{fig:SystemA03}, left panel) at the same observed wavelength (4535\AA) in both B600 spectra. Taking into account the absence of other spectral features in the R150 spectra, along with the photometric redshift of the lens galaxy, we assign the features to be \lymanalpha, which gives redshifts of $\zsource=\SystemAspeczA$ and $\SystemAspeczB$ for arcs A1 and A2, respectively. 

We estimate an Einstein radius of $\thetae \mysim \SystemAradiusA \arcsec$, which passes through both arcs, giving an enclosed mass of $\menclosed\mysim\SystemAmencAtext$.

In the DES imaging, there appear two arcs that may be counter-images. An arc north of arc A2, and extremely close to the lensing galaxy, may be a counter-image to arc A1. These objects appear more clearly in Fig.~3, panel `39' of \citet{more12}. If we assume that arc A1 has a counterpart to the northeast, we can perform a manual measurement of the image-splitting distance: for lensing systems where the source lies nearly along the line of sight through the lens, the distance between an image and its counter is related to the Einstein radius: $\Delta\theta \sim 2\thetae$. For this system, $\Delta\theta = 7.7\pm1.7\arcsec$, which gives $\thetae \sim 3.8\pm0.9\arcsec$, in agreement with the direct manual measurement of the Einstein radius.

There is yet another possible counter-image slightly north and east of arc A1. These counter-images were not targeted: for a single multi-slit mask, it was not possible to obtain spectra for these objects simultaneously with that of the brighter arcs we had chosen (see top right panel of Fig.~\ref{fig:SystemA03} for the configuration of the mask slits).

\subsection{\SystemBDES}\label{sec:sample:systemB05}
\SystemBDES\ is a newly discovered system identified in the \redmapper\ cluster sample during the targeted search. The central red lensing galaxy has a photometric redshift of $\zlens=0.841\pm0.042$, with two blue features labeled \arcA\ and \arcB\ in Fig.~\ref{fig:multipanel}b and in Fig.~\ref{fig:SystemB05} (top and middle right panels); the arcs are located east and north of the central cluster galaxy, respectively. We identify a strong emission line for \arcA\ (Fig.~\ref{fig:SystemB05}, left panel) in the GMOS R150 spectrum at an observed wavelength of 8229\AA. We associate this feature with \OII, which yields a redshift of $\zsource=\SystemBspeczA$. We do not detect any continuum, nor any emission or absorption lines, from arc \arcB.

While we do not detect features in the arc \arcB\ spectrum, its morphology suggests that it is likely to be a lensed source image. Assuming it is a feature of this system, we located the lensing mass center such that it aligns with the tangential arc B. As the circle of the Einstein radius passes through both images A and B, we use the spectroscopic redshift of arc \arcA, and we estimate a single Einstein radius of $\thetae\mysim\SystemBradiusA\arcsec$ for this system, which gives an enclosed mass of $\menclosed\mysim\SystemBmencAtext$. The source of image \arcA\ may lie outside the caustic, and thus may not be strongly lensed, only weakly magnified. In such a case, this would provide a poor estimate of the enclosed mass.

\subsection{\SystemCDES}\label{sec:sample:systemC06}
\SystemCDES\ is also a newly discovered system identified in the \redmapper\ galaxy cluster sample during the targeted search. The central red galaxy has a photometric redshift $\zlens=0.655\pm0.033$. There are two small blue arcs, \arcA\ and \arcB, to the west and north of the central red galaxy in the image, shown in Fig.~\ref{fig:multipanel}c and in Fig.~\ref{fig:SystemC06}, top and middle right panels. In the follow-up IMACS spectroscopy of the arcs, we identify emission lines at two different observed wavelengths of 6694\AA\ and 8563\AA. The GMOS B600 data for this system provide no observed features at shorter wavelengths: therefore, we identify the emission line as \OII, from which we obtain redshifts, $\zsource=\SystemCspeczA$ and $\SystemCspeczB$ for images, \arcA\ and \arcB, respectively (Fig.~\ref{fig:SystemC06}, left panel). 

Arc \arcB\ may be outside the caustic, and thus only weakly magnified and perhaps not multiply imaged. We therefore elect to use only \arcA\ (the circle for which also goes through \arcB). We then estimate a single Einstein radius of  $\thetae\mysim\SystemCradiusA\arcsec$ for \arcA, which gives an enclosed mass of $\menclosed\mysim\SystemCmencAtext$.

\subsection{\SystemDDES}\label{sec:sample:systemD07}
\SystemDDES\ was discovered with XMM-Newton---via detection of a second X-ray peak \citep{werner07}---as a cluster behind Abell 3128. 
It has also been measured in both the SPT \citep[SPT-CL J0330-5228;][]{ruel14} and Atacama Cosmology Telescope (ACT) SZ surveys \citep[ACT-CL J0330-5227;][]{menanteau10b}. The central red galaxy has a photometric redshift of $\zlens=0.463\pm0.046$. There is a very prominent blue arc (\arcA) in Fig.~\ref{fig:multipanel}d to the southwest of the central lensing galaxy. In the follow-up GMOS R150 spectroscopy of the blue arc, we identify a strong emission line at an observed wavelength of 9146\AA, which we take to be \OII, from which we obtain a redshift of $\zsource=\SystemDspeczA$ (Fig.~\ref{fig:SystemD07}). 
 
We placed the center to the northeast to coincide with the arc's curvature. We estimate an Einstein radius of $\thetae\mysim\SystemDradiusA\arcsec$ for this system. Instead of the DES photometric redshift, we used the spectroscopic redshift from \citet{werner07}, $\zsource= 0.43961\pm 0.00014$,  to estimate an enclosed mass of $\menclosed\mysim\SystemDmencAtext$. \citet{werner07} also estimates an Einstein radius of $\sim6.2\arcsec$ and an upper limit on the mass of $\sim 5\times 10^{12} \msol$: the authors noted that they only had a redshift measurement for the lens, and assumed a combined angular diameter distance of $D_{\rm L} D_{\rm S}/ D_{\rm L S} = 1\, {\rm Gpc}$ (see Eqn.~\ref{eqn:menclosed}), which corresponds to an overestimate of the source spectroscopic redshift, and thus an underestimate of the mass.

\subsection{\SystemEDES}\label{sec:sample:systemE14}
\SystemEDES\ is also a newly discovered system, identified in the \redmapper\ cluster-finder sample. Two small blue arcs, B1 and B2, lie to the northwest and southwest, respectively, of the central red galaxy; this is shown in Fig.~\ref{fig:multipanel}e and in the top and middle right panels of Fig.~\ref{fig:SystemE14}. The candidate lensing galaxy has a photometric redshift of $\zlens=0.746\pm 0.047$, in agreement with the \redmapper\ cluster photometric  redshift, $\zredmapper = 0.734\pm0.020$. 

The GMOS B600 spectroscopy reveals prominent emission lines (Fig.~\ref{fig:SystemE14}, left panel) in both arcs at the same observed wavelength of 5117\AA. Given the line strength in the B600 spectra, we expect to see other strong lines, such as $\Hbeta$ in the R150 spectrum. Therefore, without additional features, we conclude that this is \lymanalpha\ emission, from which we obtain spectroscopic redshifts of $\zsource=\SystemEspeczB$ and $\SystemEspeczC$ for B1 and B2, respectively. The object labeled \arcA\ in Fig.~\ref{fig:multipanel}e also has emission lines (not shown), which we interpret as \OII\ and \OIII. This gives a redshift of $\zsource = \SystemEspeczA$, making object \arcA\ a foreground object. 

We estimate an Einstein radius of $\thetae\mysim\SystemEradiusB\arcsec$ for this system, which gives an enclosed mass of $\menclosed\mysim\SystemEmencBtext$.

\subsection{\SystemFDES}\label{sec:sample:systemF24}
\SystemFDES\ is a galaxy cluster originally found in the Southern Cosmology Survey \citep[SCS;][]{menanteau09} and the ACT SZ Survey \citep[ACT;][]{kosowsky06, fowler07}, and it has a cluster photometric redshift of $\zcluster=0.54$ \citep{menanteau10}. It is also found in the \redmapper\ cluster-finder sample, which gives a photometric redshift $\zredmapper = 0.522\pm0.011$, in agreement with the DES photometric redshift, $\zlens=0.530\pm0.075$. 

There are two small red arcs, \arcA\ and \arcB, to the west and southeast of the central red galaxy, respectively. These are shown in Fig.~\ref{fig:multipanel}f and in the top and middle right panels of Fig.~\ref{fig:SystemF24}. The GMOS R150 spectra of these two arcs show prominent emission lines at two different observed wavelengths---8023\AA\ for \arcA\ and 7070\AA\ for \arcB---which we identify as \OII. These yield source redshifts of $\zsource=\SystemFspeczA$ and $\SystemFspeczB$ for \arcA\ and \arcB, respectively (Fig.~\ref{fig:SystemF24}, left panel).

We estimate an Einstein radius of $\thetae\mysim\SystemFradiusA\arcsec$ for \arcA\ and $\thetae\mysim\SystemFradiusB\arcsec$ for \arcB, which give enclosed masses of $\menclosed\mysim\SystemFmencAtext$ and $\menclosed\mysim\SystemFmencBtext$, respectively. 

We note that normally one would expect the lower redshift source to have a smaller $\thetae$ than the higher redshift one, but that does not seem to be the case here. Arc \arcB\ may be outside the caustic, and thus only weakly magnified and not multiply imaged; it is also not clearly tangentially aligned with the candidate lens, which makes it less likely to be a lensed source. 

There also exists a faint image to the west of arc \arcB, almost directly south of the central red lensing galaxy, that may be another source or counter-image. To provide sufficient sky coverage in the slit for arc \arcB, this  image could not also be observed in a single mask.

\section{Discussion}\label{sec:discussion}

Many of our unconfirmed systems exhibit clear lensing features but have no detectable continuum or spectral features, with integration times of about $1$ hour on an $8$m-class telescope\footnote{For a number of the targets, we did not obtain the 3.7-hour GMOS R150 observations, which might have helped to confirm more systems.}. The need for long integration time highlights the challenge of obtaining spectra for faint objects found by deep optical surveys, like DES: the challenge will only be greater for the Large Synoptic Survey Telescope\footnote{\url{http://www.lsst.org/lsst/}} \citep[LSST;][]{LSST09}. One faces the prospect of long integration times, which limits the number of systems that can be followed up with a typical program. For systems that exhibit a clear lensing morphology and that have detectable continuum emission but no visible features, IR data may be necessary to confirm them. 

A number of large telescopes housing UV to infrared spectroscopic instruments are planned for the coming 10-20 years, which will help to alleviate this pressure. The Thirty Meter Telescope\footnote{\url{http://ast.noao.edu/system/us-tmt-liaison/survey-faq}} will require a mere hour of integration on $i\sim 25$ for low-noise spectroscopic measurements---similarly for the Giant Magellan Telescope\footnote{\url{http://www.gmto.org/}}. However, with their expected high subscription rate, it remains to be seen if there will be enough time to follow up the systems predicted for projects like LSST.

While Gemini's GMOS instrument covers optical wavelengths multi-slit or long-slit mode, the Very Large Telescope's X-shooter\footnote{\url{https://www.eso.org/sci/facilities/paranal/instruments/xshooter.html}} provides long-slit capability for strong lens follow-up simultaneously in optical and NIR wavelengths. X-shooter is the first 2nd-generation instrument on the 8m VLT: it has an echelle spectrograph that 1) covers a spectral range $300-2500$nm at intermediate resolution (R$\sim 4,000 − 17,000$) and 2) operates in long-slit or IFU mode\citep{vernet11}. In addition new instruments are being proposed for the Gemini Observatories via the Gemini Instrument Feasibility study\footnote{\url{https://www.gemini.edu/node/12362}}. For example, the Gemini Multi-Obect eXtra-wide-band Spectrograph \citep[GMOX;][]{robberto15} has a three-arm spectrograph covering the entire optical and near-IR spectrum that is available from the ground. Another example, OCTOCAM, is a multi-channel imager and long-slit spectrograph, also covering the optical and near-IR wavelengths \citep{deugarte15}. Both GMOX and OCTOCAM add integral field capability, which are ideal for extended lenses. If integration times are sufficiently low for these new instruments, the Gemini observatories may provide a competitive resource for follow-up of lensing systems in future surveys.

\section{Summary}\label{sec:summary}

In this paper, we presented the first confirmed group- and cluster-scale strong lensing systems found in DES data. These systems were discovered through visual identification using both targeted and \opensearch\ searches. We confirmed these systems with a combination of spectroscopy from GMOS on the Gemini South telescope and IMACS on the Magellan/Baade telescope.

Two of these systems are of particular interest, because they have source galaxies at redshifts  $z=2.7251$ and $z=3.22086$. Due to the magnification provided by lensing, these sources are typically among the brightest galaxies in their redshift ranges and can provide an important window into the star formation history and galaxy formation at these cosmic epochs. 

The candidate lensing systems discovered in DES SV data span a range of arc sizes, much like the currently known sample of candidates from searches during the last decade. Due to the seeing limitations of DES (median expected seeing $\sim 0.9\arcsec $, systems with smaller arc separations were more difficult to identify, especially during the non-targeted search. Thus, our visual inspection methods likely yielded higher completeness in the targeted search of  \redmapper\ and SPT clusters.

\tablelensingfeatures

As we examine more of the DES wide-field area, we expect to find many more systems. Through comparison with the CFHTLS data, we expect there to be $\sim1700$ lenses in the full survey, which is similar to the number found in the preceding decade of survey searches in optical and IR wavelengths listed in \S\ref{sec:introduction}. Independently, \citet{collett15} predicts 800 galaxy-scale lenses.

The Year 1 DES data cover an area of $\sim1800$ \sqdeg, and we have already conducted targeted searches, identifying a number of promising systems for follow-up. We have also expanded the search techniques to use arc-finding  \citep[e.g.,][]{more12} and ring-finding algorithms \citep[e.g.,][]{gavazzi14}, as well as catalog-based photometry searches similar to those employed in the SBAS \citep[e.g.,][]{diehl09}. We continue the dedicated program, STRIDES \citep{treu15}, to search for lensed quasars. With these tools, our searches through Year 1 data will be more effective than in SV for finding galaxy-scale lenses. The high yields of DES and future surveys will likely pose a challenge for spectroscopic follow-up, and thus a challenge to maximize strong lensing for use in constraining cosmology.

\section*{Acknowledgments}

We performed this analysis under research operated by Fermi Research Alliance, LLC under Contract No. De-AC02-07CH11359 with the United States Department of Energy.

We are grateful for the extraordinary contributions of our CTIO colleagues and the DES Camera, Commissioning and Science Verification teams for achieving excellent instrument and telescope conditions that have made this work possible. The success of this project also relies critically on the expertise and dedication of the DES Data Management organization.

Funding for the DES Projects has been provided by the U.S. Department of Energy, the U.S. National Science Foundation, the Ministry of Science and Education of Spain, the Science and Technology Facilities Council of the United Kingdom, the Higher Education Funding Council for England, the National Center for Supercomputing Applications at the University of Illinois at Urbana-Champaign, the Kavli Institute of Cosmological Physics at the University of Chicago, the Center for Cosmology and Astro-Particle Physics at the Ohio State University, the Mitchell Institute for Fundamental Physics and Astronomy at Texas A\&M University, Financiadora de Estudos e Projetos, Funda\c{c}\~{a}o Carlos Chagas Filho de Amparo \`{a} Pesquisa do Estado do Rio de Janeiro, Conselho Nacional de Desenvolvimento Científico e Tecnol\'{o}gico and the Minist\'{e}rio da Ci\^{e}ncia e Tecnologia, the Deutsche Forschungsgemeinschaft and the Collaborating Institutions in the Dark Energy Survey. The DES data management system is supported by the National Science Foundation under Grant Number AST-1138766. The DES participants from Spanish institutions are partially supported by MINECO under grants AYA2012-39559, ESP2013-48274, FPA2013-47986, and Centro de Excelencia Severo Ochoa SEV-2012-0234, some of which include ERDF funds from the European Union.

The Collaborating Institutions are Argonne National Laboratory, the University of California at Santa Cruz, the University of Cambridge, Centro de Investigaciones Energ\'{e}ticas, Medioambientales y Tecnol\'{o}gicas-Madrid, the University of Chicago, University College London, the DES-Brazil Consortium, the Eidgenoessische Technische Hochschule (ETH) Zurich, Fermi National Accelerator Laboratory, the University of Edinburgh, the University of Illinois at Urbana-Champaign, the Institut de Ci\`{e}ncies de l'Espai (IEEC/CSIC), the Institut de F\'{i}sica d'Altes Energies, Lawrence Berkeley National Laboratory, the Ludwig-Maximilians Universit\"{a}t and the associated Excellence Cluster Universe, the University of Michigan, the National Optical Astronomy Observatory, the University of Nottingham, the Ohio State University, the University of Pennsylvania, the University of Portsmouth, SLAC National Accelerator Laboratory, Stanford University, the University of Sussex, and Texas A\&M University.

We are also grateful to Janani Sivakumar, student at the Illinois Mathematics and Science Academy, as well as Caroline Odden and her students, Kathryn Chapman, George Avecillas and Matthew Simon, at the Phillips Academy for their contributions to the scanning effort. We also thank Matthew Becker and Risa Wechsler for helpful discussions on the topic of galaxy cluster richnesses and masses. 

This work is based in part on observations obtained at the Gemini Observatory, which is operated by the Association of Universities for Research in Astronomy, Inc., under a cooperative agreement with the NSF on behalf of the Gemini partnership: the National Science Foundation (United States), the National Research Council (Canada), CONICYT (Chile), the Australian Research Council (Australia), Minist\'{e}rio da Ci\^{e}ncia, Tecnologia e Inova\c{c}\~{a}o (Brazil) and Ministerio de Ciencia, Tecnolog\'{i}a e Innovaci\'{o}n Productiva (Argentina). The data was acquired through the Gemini Science Archive and processed using the Gemini IRAF package v2.16.

Thanks to the entire Las Campanas Observatory staff for their help in the acquisition of the Magellan/IMACS data \citep[Inamori Magellan Areal Camera and Spectrograph;][]{bigelow03}. This paper includes data gathered with the 6.5m Magellan Telescopes located at Las Campanas Observatory, Chile.

C. Furlanetto acknowledges funding from CAPES (proc. 12203-1). This paper has gone through internal review by the DES collaboration. This research has made use of NASA's Astrophysics Data System.

C. De Bom would like to thank CNPq for the financial support.

\bibliography{DESStrongLensPaper}

\begin{thebibliography}{}
\expandafter\ifx\csname natexlab\endcsname\relax\def\natexlab#1{#1}\fi

\bibitem[{{Agnello} {et~al.}(2015){Agnello}, {Treu}, {Ostrovski}, {Schechter},
  {Buckley-Geer}, {Lin}, {Auger}, {Courbin}, {Fassnacht}, {Frieman},
  {Kuropatkin}, {Marshall}, {McMahon}, {Meylan}, {More}, {Suyu}, {Rusu},
  {Finley}, {Abbott}, {Abdalla}, {Allam}, {Annis}, {Banerji},
  {Benoit-L{\'e}vy}, {Bertin}, {Brooks}, {Burke}, {Rosell}, {Kind},
  {Carretero}, {Cunha}, {D'Andrea}, {da Costa}, {Desai}, {Diehl}, {Dietrich},
  {Doel}, {Eifler}, {Estrada}, {Neto}, {Flaugher}, {Fosalba}, {Gerdes},
  {Gruen}, {Gutierrez}, {Honscheid}, {James}, {Kuehn}, {Lahav}, {Lima}, {Maia},
  {March}, {Marshall}, {Martini}, {Melchior}, {Miller}, {Miquel}, {Nichol},
  {Ogando}, {Plazas}, {Reil}, {Romer}, {Roodman}, {Sako}, {Sanchez},
  {Santiago}, {Scarpine}, {Schubnell}, {Sevilla-Noarbe}, {Smith},
  {Soares-Santos}, {Sobreira}, {Suchyta}, {Swanson}, {Tarle}, {Thaler},
  {Tucker}, {Walker}, {Wechsler}, \& {Zhang}}]{agnello15}
{Agnello}, A., {Treu}, T., {Ostrovski}, F., {et~al.} 2015, \mnras, 454, 1260

\bibitem[{{Allam} {et~al.}(2007){Allam}, {Tucker}, {Lin}, {Diehl}, {Annis},
  {Buckley-Geer}, \& {Frieman}}]{allam07}
{Allam}, S.~S., {Tucker}, D.~L., {Lin}, H., {et~al.} 2007, \apjl, 662, L51

\bibitem[{{ALMA Partnership} {et~al.}(2015)}]{alma15}
{ALMA Partnership}, {et~al.} 2015, \apjl, 808, L4

\bibitem[{{Belokurov} {et~al.}(2009){Belokurov}, {Evans}, {Hewett}, {Moiseev},
  {McMahon}, {Sanchez}, \& {King}}]{belokurov09}
{Belokurov}, V., {Evans}, N.~W., {Hewett}, P.~C., {et~al.} 2009, \mnras, 392,
  104

\bibitem[{{Belokurov} {et~al.}(2007)}]{belokurov07}
{Belokurov}, V., {et~al.} 2007, \apjl, 671, L9

\bibitem[{{Bertin}(2011)}]{bertin11}
{Bertin}, E. 2011, in Astronomical Society of the Pacific Conference Series,
  Vol. 442, Astronomical Data Analysis Software and Systems XX, ed. I.~N.
  {Evans}, A.~{Accomazzi}, D.~J. {Mink}, \& A.~H. {Rots}, 435

\bibitem[{{Bertin} \& {Arnouts}(1996)}]{bertin96}
{Bertin}, E., \& {Arnouts}, S. 1996, \aaps, 117, 393

\bibitem[{{Bigelow} \& {Dressler}(2003)}]{bigelow03}
{Bigelow}, B.~C., \& {Dressler}, A.~M. 2003, in Society of Photo-Optical
  Instrumentation Engineers (SPIE) Conference Series, Vol. 4841, Instrument
  Design and Performance for Optical/Infrared Ground-based Telescopes, ed.
  M.~{Iye} \& A.~F.~M. {Moorwood}, 1727--1738

\bibitem[{{Blain}(1996)}]{blain96}
{Blain}, A.~W. 1996, \mnras, 283, 1340

\bibitem[{{Bleem} {et~al.}(2015)}]{bleem15}
{Bleem}, L.~E., {et~al.} 2015, \apjs, 216, 27

\bibitem[{{B{\"o}hringer} {et~al.}(2004)}]{bohringer04}
{B{\"o}hringer}, H., {et~al.} 2004, \aap, 425, 367

\bibitem[{{Bolton} {et~al.}(2008){Bolton}, {Burles}, {Koopmans}, {Treu},
  {Gavazzi}, {Moustakas}, {Wayth}, \& {Schlegel}}]{bolton08}
{Bolton}, A.~S., {Burles}, S., {Koopmans}, L.~V.~E., {et~al.} 2008, \apj, 682,
  964

\bibitem[{{Browne} {et~al.}(2003){Browne}, {Wilkinson}, {Jackson}, {Myers},
  {Fassnacht}, {Koopmans}, {Marlow}, {Norbury}, {Rusin}, {Sykes}, {Biggs},
  {Blandford}, {de Bruyn}, {Chae}, {Helbig}, {King}, {McKean}, {Pearson},
  {Phillips}, {Readhead}, {Xanthopoulos}, \& {York}}]{browne03}
{Browne}, I.~W.~A., {Wilkinson}, P.~N., {Jackson}, N.~J.~F., {et~al.} 2003,
  \mnras, 341, 13

\bibitem[{{Buckley-Geer} {et~al.}(2011)}]{buckleygeer11}
{Buckley-Geer}, E.~J., {et~al.} 2011, \apj, 742, 48

\bibitem[{{Bussmann} {et~al.}(2013)}]{bussmann13}
{Bussmann}, R.~S., {et~al.} 2013, \apj, 779, 25

\bibitem[{{Cabanac} {et~al.}(2007)}]{cabanac07}
{Cabanac}, R.~A., {et~al.} 2007, \aap, 461, 813

\bibitem[{{Carlstrom} {et~al.}(2011)}]{carlstrom11}
{Carlstrom}, J.~E., {et~al.} 2011, \pasp, 123, 568

\bibitem[{{Collett}(2015)}]{collett15}
{Collett}, T.~E. 2015, \apj, 811, 20

\bibitem[{{Collett} \& {Auger}(2014)}]{collett14}
{Collett}, T.~E., \& {Auger}, M.~W. 2014, \mnras, 443, 969

\bibitem[{{Collett} {et~al.}(2012){Collett}, {Auger}, {Belokurov}, {Marshall},
  \& {Hall}}]{collett12}
{Collett}, T.~E., {Auger}, M.~W., {Belokurov}, V., {Marshall}, P.~J., \&
  {Hall}, A.~C. 2012, \mnras, 424, 2864

\bibitem[{{de Ugarte Postigo}(2015)}]{deugarte15}
{de Ugarte Postigo}, A. 2015, IAU General Assembly, 22, 57336

\bibitem[{{Desai} {et~al.}(2012)}]{desai12}
{Desai}, S., {et~al.} 2012, \apj, 757, 83

\bibitem[{{Diehl} {et~al.}(2009)}]{diehl09}
{Diehl}, H.~T., {et~al.} 2009, \apj, 707, 686

\bibitem[{Diehl {et~al.}(2014)Diehl, Abbott, Annis, Armstrong, Baruah, Bermeo,
  Bernstein, Beynon, Bruderer, Buckley-Geer, Campbell, Capozzi, Carter, Casas,
  Clerkin, Covarrubias, Cuhna, D'Andrea, da~Costa, Das, DePoy, Dietrich,
  Drlica-Wagner, Elliott, Eifler, Estrada, Etherington, Flaugher, Frieman,
  Fausti~Neto, Gelman, Gerdes, Gruen, Gruendl, Hao, Head, Helsby, Hoffman,
  Honscheid, James, Johnson, Kacprzac, Katsaros, Kennedy, Kent, Kessler, Kim,
  Krause, Kron, Kuhlmann, Kunder, Li, Lin, MacCrann, March, Marshall, Neilsen,
  Nugent, Martini, Melchior, Menanteau, Nichol, Nord, Ogando, Old,
  Papadopoulos, Patton, Petravick, Plazas, Poulton, Pujol, Reil, Rigby, Romer,
  Roodman, Rooney, Sanchez~Alvaro, Serrano, Sheldon, Smith, Smith,
  Soares-Santos, Soumagnac, Spinka, Suchyta, Tucker, Walker, Wester, Wiesner,
  Wilcox, Williams, Yanny, \& Zhang}]{diehl14}
Diehl, H.~T., Abbott, T. M.~C., Annis, J., {et~al.} 2014, The Dark Energy
  Survey and operations: Year 1, doi:10.1117/12.2056982

\bibitem[{{Dressler} {et~al.}(2011)}]{dressler11}
{Dressler}, A., {et~al.} 2011, \pasp, 123, 288

\bibitem[{{Egami} {et~al.}(2010){Egami}, {Rex}, {Rawle},
  {P{\'e}rez-Gonz{\'a}lez}, {Richard}, {Kneib}, {Schaerer}, {Altieri},
  {Valtchanov}, {Blain}, {Fadda}, {Zemcov}, {Bock}, {Boone}, {Bridge},
  {Clement}, {Combes}, {Dessauges-Zavadsky}, {Dowell}, {Ilbert}, {Ivison},
  {Jauzac}, {Lutz}, {Metcalfe}, {Omont}, {Pell{\'o}}, {Pereira}, {Rieke},
  {Rodighiero}, {Smail}, {Smith}, {Tramoy}, {Walth}, {van der Werf}, \&
  {Werner}}]{egami10}
{Egami}, E., {Rex}, M., {Rawle}, T.~D., {et~al.} 2010, \aap, 518, L12

\bibitem[{{Fassnacht} {et~al.}(1999)}]{fassnacht99}
{Fassnacht}, C.~D., {et~al.} 1999, \aj, 117, 658

\bibitem[{{Faure} {et~al.}(2008)}]{faure08}
{Faure}, C., {et~al.} 2008, \apjs, 176, 19

\bibitem[{{Flaugher} {et~al.}(2015)}]{flaugher15}
{Flaugher}, B., {et~al.} 2015, ArXiv e-prints, arXiv:1504.02900

\bibitem[{{Fowler} {et~al.}(2007){Fowler}, {Niemack}, {Dicker}, {Aboobaker},
  {Ade}, {Battistelli}, {Devlin}, {Fisher}, {Halpern}, {Hargrave}, {Hincks},
  {Kaul}, {Klein}, {Lau}, {Limon}, {Marriage}, {Mauskopf}, {Page}, {Staggs},
  {Swetz}, {Switzer}, {Thornton}, \& {Tucker}}]{fowler07}
{Fowler}, J.~W., {Niemack}, M.~D., {Dicker}, S.~R., {et~al.} 2007, \ao, 46,
  3444

\bibitem[{GAIA(2015)}]{gaiawebsite}
GAIA. 2015, Graphical Astronomy and Image Analysis Tool,
  \url{http://star-www.dur.ac.uk/~pdraper/gaia/gaia.html}

\bibitem[{{Gavazzi} {et~al.}(2014){Gavazzi}, {Marshall}, {Treu}, \&
  {Sonnenfeld}}]{gavazzi14}
{Gavazzi}, R., {Marshall}, P.~J., {Treu}, T., \& {Sonnenfeld}, A. 2014, \apj,
  785, 144

\bibitem[{{Gavazzi} {et~al.}(2008){Gavazzi}, {Treu}, {Koopmans}, {Bolton},
  {Moustakas}, {Burles}, \& {Marshall}}]{gavazzi08}
{Gavazzi}, R., {Treu}, T., {Koopmans}, L.~V.~E., {et~al.} 2008, \apj, 677, 1046

\bibitem[{Gruendl {et~al.}(2015, in prep.)}]{gruendl15}
Gruendl, R., {et~al.} 2015, in prep.

\bibitem[{{Hennawi} {et~al.}(2008)}]{hennawi08}
{Hennawi}, J.~F., {et~al.} 2008, \aj, 135, 664

\bibitem[{{Hook} {et~al.}(2004){Hook}, {J{\o}rgensen}, {Allington-Smith},
  {Davies}, {Metcalfe}, {Murowinski}, \& {Crampton}}]{2004PASP..116..425H}
{Hook}, I.~M., {J{\o}rgensen}, I., {Allington-Smith}, J.~R., {et~al.} 2004,
  \pasp, 116, 425

\bibitem[{{Horesh} {et~al.}(2005){Horesh}, {Ofek}, {Maoz}, {Bartelmann},
  {Meneghetti}, \& {Rix}}]{horesh05}
{Horesh}, A., {Ofek}, E.~O., {Maoz}, D., {et~al.} 2005, \apj, 633, 768

\bibitem[{{Jackson}(2008)}]{jackson08}
{Jackson}, N. 2008, \mnras, 389, 1311

\bibitem[{{Joseph} {et~al.}(2014){Joseph}, {Courbin}, {Metcalf}, {Giocoli},
  {Hartley}, {Jackson}, {Bellagamba}, {Kneib}, {Koopmans}, {Lemson},
  {Meneghetti}, {Meylan}, {Petkova}, \& {Pires}}]{joseph14}
{Joseph}, R., {Courbin}, F., {Metcalf}, R.~B., {et~al.} 2014, \aap, 566, A63

\bibitem[{{Jullo} {et~al.}(2010){Jullo}, {Natarajan}, {Kneib}, {D'Aloisio},
  {Limousin}, {Richard}, \& {Schimd}}]{jullo10}
{Jullo}, E., {Natarajan}, P., {Kneib}, J.-P., {et~al.} 2010, Science, 329, 924

\bibitem[{{Kosowsky}(2006)}]{kosowsky06}
{Kosowsky}, A. 2006, 50, 969

\bibitem[{{Kubik}(2007)}]{kubik07}
{Kubik}, D. 2007, Master's thesis, Northern Illinois University

\bibitem[{{Kubo} {et~al.}(2009){Kubo}, {Allam}, {Annis}, {Buckley-Geer},
  {Diehl}, {Kubik}, {Lin}, \& {Tucker}}]{kubo09}
{Kubo}, J.~M., {Allam}, S.~S., {Annis}, J., {et~al.} 2009, \apjl, 696, L61

\bibitem[{{Kubo} {et~al.}(2010)}]{kubo10}
{Kubo}, J.~M., {et~al.} 2010, \apjl, 724, L137

\bibitem[{{Kurtz} \& {Mink}(1998)}]{kurtz98}
{Kurtz}, M.~J., \& {Mink}, D.~J. 1998, \pasp, 110, 934

\bibitem[{{Lenzen} {et~al.}(2004){Lenzen}, {Schindler}, \&
  {Scherzer}}]{lenzen04}
{Lenzen}, F., {Schindler}, S., \& {Scherzer}, O. 2004, \aap, 416, 391

\bibitem[{{Lin} {et~al.}(2009)}]{lin09}
{Lin}, H., {et~al.} 2009, \apj, 699, 1242

\bibitem[{{LSST Science Collaboration} {et~al.}(2009){LSST Science
  Collaboration}, {Abell}, {Allison}, {Anderson}, {Andrew}, {Angel}, {Armus},
  {Arnett}, {Asztalos}, {Axelrod}, \& et~al.}]{LSST09}
{LSST Science Collaboration}, {Abell}, P.~A., {Allison}, J., {et~al.} 2009,
  ArXiv e-prints, arXiv:0912.0201

\bibitem[{{McKean} {et~al.}(2005)}]{mckean05}
{McKean}, J.~P., {et~al.} 2005, \mnras, 356, 1009

\bibitem[{{Menanteau} {et~al.}(2009){Menanteau}, {Hughes}, {Jimenez},
  {Hernandez-Monteagudo}, {Verde}, {Kosowsky}, {Moodley}, {Infante}, \&
  {Roche}}]{menanteau09}
{Menanteau}, F., {Hughes}, J.~P., {Jimenez}, R., {et~al.} 2009, \apj, 698, 1221

\bibitem[{{Menanteau} {et~al.}(2010{\natexlab{a}})}]{menanteau10}
{Menanteau}, F., {et~al.} 2010{\natexlab{a}}, \apjs, 191, 340

\bibitem[{{Menanteau} {et~al.}(2010{\natexlab{b}})}]{menanteau10b}
---. 2010{\natexlab{b}}, \apj, 723, 1523

\bibitem[{{Menanteau} {et~al.}(2012)}]{menanteau12}
---. 2012, \apj, 748, 7

\bibitem[{{More} {et~al.}(2012)}]{more12}
{More}, A., {et~al.} 2012, \apj, 749, 38

\bibitem[{{More} {et~al.}(2016){More}, {Verma}, {Marshall}, {More}, {Baeten},
  {Wilcox}, {Macmillan}, {Cornen}, {Kapadia}, {Parrish}, {Snyder}, {Davis},
  {Gavazzi}, {Lintott}, {Simpson}, {Miller}, {Smith}, {Paget}, {Saha},
  {K{\"u}ng}, \& {Collett}}]{more16}
{More}, A., {Verma}, A., {Marshall}, P.~J., {et~al.} 2016, \mnras, 455, 1191

\bibitem[{{Myers} {et~al.}(2003){Myers}, {Jackson}, {Browne}, {de Bruyn},
  {Pearson}, {Readhead}, {Wilkinson}, {Biggs}, {Blandford}, {Fassnacht},
  {Koopmans}, {Marlow}, {McKean}, {Norbury}, {Phillips}, {Rusin}, {Shepherd},
  \& {Sykes}}]{myers03}
{Myers}, S.~T., {Jackson}, N.~J., {Browne}, I.~W.~A., {et~al.} 2003, \mnras,
  341, 1

\bibitem[{{Narayan} \& {Bartelmann}(1996)}]{narayan96}
{Narayan}, R., \& {Bartelmann}, M. 1996, ArXiv Astrophysics e-prints,
  astro-ph/9606001

\bibitem[{{Negrello} {et~al.}(2007){Negrello}, {Perrotta},
  {Gonz{\'a}lez-Nuevo}, {Silva}, {de Zotti}, {Granato}, {Baccigalupi}, \&
  {Danese}}]{negrello07}
{Negrello}, M., {Perrotta}, F., {Gonz{\'a}lez-Nuevo}, J., {et~al.} 2007,
  \mnras, 377, 1557

\bibitem[{{Negrello} {et~al.}(2010)}]{negrello10}
{Negrello}, M., {et~al.} 2010, Science, 330, 800

\bibitem[{{Oguri} \& {Marshall}(2010)}]{oguri10}
{Oguri}, M., \& {Marshall}, P.~J. 2010, \mnras, 405, 2579

\bibitem[{{Planck Collaboration} {et~al.}(2015){Planck Collaboration}, {Ade},
  {Aghanim}, {Arnaud}, {Ashdown}, {Aumont}, {Baccigalupi}, {Banday},
  {Barreiro}, {Bartlett}, \& et~al.}]{planckcosmo15}
{Planck Collaboration}, {Ade}, P.~A.~R., {Aghanim}, N., {et~al.} 2015, ArXiv
  e-prints, arXiv:1502.01589

\bibitem[{{Reblinsky} \& {Bartelmann}(1999)}]{reblinsky99}
{Reblinsky}, K., \& {Bartelmann}, M. 1999, \aap, 345, 1

\bibitem[{{Refsdal}(1964)}]{refsdal64}
{Refsdal}, S. 1964, \mnras, 128, 307

\bibitem[{{Robberto} {et~al.}(2015){Robberto}, {Heckman}, {Gennaro}, {Deustua},
  {MacKenty}, {Ninkov}, {Becker}, {Bianchi}, {Bellini}, {Calamida}, {Kalirai},
  {Lotz}, {Sabbi}, {Tumlinson}, {Smee}, \& {Barkhouser}}]{robberto15}
{Robberto}, M., {Heckman}, T., {Gennaro}, M., {et~al.} 2015, IAU General
  Assembly, 22, 57947

\bibitem[{{Ruel} {et~al.}(2014)}]{ruel14}
{Ruel}, J., {et~al.} 2014, \apj, 792, 45

\bibitem[{{Rykoff} {et~al.}(2014)}]{rykoff14}
{Rykoff}, E.~S., {et~al.} 2014, \apj, 785, 104

\bibitem[{{S{\'a}nchez} {et~al.}(2014)}]{sanchez14}
{S{\'a}nchez}, C., {et~al.} 2014, \mnras, 445, 1482

\bibitem[{{Saro} {et~al.}(2015){Saro}, , {et~al.}}]{saro15}
{Saro}, A., , {et~al.} 2015, \mnras, 454, 2305

\bibitem[{{Seidel} \& {Bartelmann}(2007)}]{seidel07}
{Seidel}, G., \& {Bartelmann}, M. 2007, \aap, 472, 341

\bibitem[{{Smail} {et~al.}(1997){Smail}, {Ivison}, \& {Blain}}]{smail97}
{Smail}, I., {Ivison}, R.~J., \& {Blain}, A.~W. 1997, \apjl, 490, L5

\bibitem[{{Suyu} {et~al.}(2013)}]{suyu13}
{Suyu}, S.~H., {et~al.} 2013, \apj, 766, 70

\bibitem[{{Swinbank} {et~al.}(2010)}]{swinbank10}
{Swinbank}, A.~M., {et~al.} 2010, \mnras, 405, 234

\bibitem[{{Treu}(2010)}]{treu10}
{Treu}, T. 2010, \araa, 48, 87

\bibitem[{{Treu} {et~al.}(2015){Treu}, {Agnello}, \& {Strides Team}}]{treu15}
{Treu}, T., {Agnello}, A., \& {Strides Team}. 2015, in American Astronomical
  Society Meeting Abstracts, Vol. 225, American Astronomical Society Meeting
  Abstracts, \#318.04

\bibitem[{{Vernet} {et~al.}(2011)}]{vernet11}
{Vernet}, J., {et~al.} 2011, \aap, 536, A105

\bibitem[{{Vieira} {et~al.}(2010)}]{vieira10}
{Vieira}, J.~D., {et~al.} 2010, \apj, 719, 763

\bibitem[{{Vieira} {et~al.}(2013)}]{vieira13}
---. 2013, \nat, 495, 344

\bibitem[{{Werner} {et~al.}(2007){Werner}, {Churazov}, {Finoguenov},
  {Markevitch}, {Burenin}, {Kaastra}, \& {B{\"o}hringer}}]{werner07}
{Werner}, N., {Churazov}, E., {Finoguenov}, A., {et~al.} 2007, \aap, 474, 707

\bibitem[{{Wiesner} {et~al.}(2012){Wiesner}, {Lin}, {Allam}, {Annis},
  {Buckley-Geer}, {Diehl}, {Kubik}, {Kubo}, \& {Tucker}}]{wiesner12}
{Wiesner}, M.~P., {Lin}, H., {Allam}, S.~S., {et~al.} 2012, \apj, 761, 1

\bibitem[{{Willis} {et~al.}(2006){Willis}, {Hewett}, {Warren}, {Dye}, \&
  {Maddox}}]{willis06}
{Willis}, J.~P., {Hewett}, P.~C., {Warren}, S.~J., {Dye}, S., \& {Maddox}, N.
  2006, \mnras, 369, 1521

\bibitem[{{York} {et~al.}(2000){York}, {Adelman}, {Anderson}, {Anderson},
  {Annis}, {Bahcall}, {Bakken}, {Barkhouser}, {Bastian}, {Berman}, {Boroski},
  {Bracker}, {Briegel}, {Briggs}, {Brinkmann}, {Brunner}, {Burles}, {Carey},
  {Carr}, {Castander}, {Chen}, {Colestock}, {Connolly}, {Crocker}, {Csabai},
  {Czarapata}, {Davis}, {Doi}, {Dombeck}, {Eisenstein}, {Ellman}, {Elms},
  {Evans}, {Fan}, {Federwitz}, {Fiscelli}, {Friedman}, {Frieman}, {Fukugita},
  {Gillespie}, {Gunn}, {Gurbani}, {de Haas}, {Haldeman}, {Harris}, {Hayes},
  {Heckman}, {Hennessy}, {Hindsley}, {Holm}, {Holmgren}, {Huang}, {Hull},
  {Husby}, {Ichikawa}, {Ichikawa}, {Ivezi{\'c}}, {Kent}, {Kim}, {Kinney},
  {Klaene}, {Kleinman}, {Kleinman}, {Knapp}, {Korienek}, {Kron}, {Kunszt},
  {Lamb}, {Lee}, {Leger}, {Limmongkol}, {Lindenmeyer}, {Long}, {Loomis},
  {Loveday}, {Lucinio}, {Lupton}, {MacKinnon}, {Mannery}, {Mantsch}, {Margon},
  {McGehee}, {McKay}, {Meiksin}, {Merelli}, {Monet}, {Munn}, {Narayanan},
  {Nash}, {Neilsen}, {Neswold}, {Newberg}, {Nichol}, {Nicinski}, {Nonino},
  {Okada}, {Okamura}, {Ostriker}, {Owen}, {Pauls}, {Peoples}, {Peterson},
  {Petravick}, {Pier}, {Pope}, {Pordes}, {Prosapio}, {Rechenmacher}, {Quinn},
  {Richards}, {Richmond}, {Rivetta}, {Rockosi}, {Ruthmansdorfer}, {Sandford},
  {Schlegel}, {Schneider}, {Sekiguchi}, {Sergey}, {Shimasaku}, {Siegmund},
  {Smee}, {Smith}, {Snedden}, {Stone}, {Stoughton}, {Strauss}, {Stubbs},
  {SubbaRao}, {Szalay}, {Szapudi}, {Szokoly}, {Thakar}, {Tremonti}, {Tucker},
  {Uomoto}, {Vanden Berk}, {Vogeley}, {Waddell}, {Wang}, {Watanabe},
  {Weinberg}, {Yanny}, {Yasuda}, \& {SDSS Collaboration}}]{york00}
{York}, D.~G., {Adelman}, J., {Anderson}, Jr., J.~E., {et~al.} 2000, \aj, 120,
  1579

\bibitem[{{York} {et~al.}(2005){York}, {Jackson}, {Browne}, {Koopmans},
  {McKean}, {Norbury}, {Biggs}, {Blandford}, {de Bruyn}, {Fassnacht}, {Myers},
  {Pearson}, {Phillips}, {Readhead}, {Rusin}, \& {Wilkinson}}]{york05}
{York}, T., {Jackson}, N., {Browne}, I.~W.~A., {et~al.} 2005, \mnras, 361, 259

\end{thebibliography}
\bibliographystyle{apj}

\FigureSystemA

\FigureSystemB

\FigureSystemC

\FigureSystemD

\FiguresystemE

\FigureSystemF

\tableredmapper

\end{document}